\documentclass[prd,amsmath,amssymb,showpacs,superscriptaddress,nofootinbib,twocolumn]{revtex4-1}
\usepackage{arydshln}
\usepackage{lineno}
\usepackage{tikz}
\usepackage{float}
\usepackage{graphicx}
\usepackage{dcolumn}
\usepackage{bm}
\usepackage{rotating}
\usepackage{color}
\usepackage{subfigure}
\usepackage{pstricks}
\usepackage{soul}
\usepackage{overpic}
\usepackage[english]{babel}
\usepackage[colorlinks,urlcolor=blue,linkcolor=blue,anchorcolor=blue,citecolor=blue]{hyperref}
\usepackage{colortbl}
\usepackage{multirow}
\usepackage{amsmath}
\usepackage{booktabs}
\usepackage{rotating}
\newcommand{\PreserveBackslash}[1]{\let\temp=\\#1\let\\=\temp}
\newcolumntype{C}[1]{>{\PreserveBackslash\centering}p{#1}}
\newcolumntype{R}[1]{>{\PreserveBackslash\raggedleft}p{#1}}
\newcolumntype{L}[1]{>{\PreserveBackslash\raggedright}p{#1}}

\newcommand{\kshort}{K^0_S}

\begin{document}

\title{\boldmath Amplitude analysis and branching fraction measurement of $D_s^+ \to K^-K^+\pi^+\pi^0$}

\author{
\begin{small}
\begin{center}
M.~Ablikim$^{1}$, M.~N.~Achasov$^{10,c}$, P.~Adlarson$^{67}$, S.~Ahmed$^{15}$, M.~Albrecht$^{4}$, R.~Aliberti$^{28}$, A.~Amoroso$^{66A,66C}$, M.~R.~An$^{32}$, Q.~An$^{63,49}$, X.~H.~Bai$^{57}$
% ...
\\
\vspace{0.2cm}
(BESIII Collaboration)\\
\vspace{0.2cm} {\it
$^{1}$ Institute of High Energy Physics, Beijing 100049, People's Republic of China\\
% ...
$^{4}$ Bochum Ruhr-University, D-44780 Bochum, Germany\\
% ...
$^{10}$ G.I. Budker Institute of Nuclear Physics SB RAS (BINP), Novosibirsk 630090, Russia\\
% ...
$^{15}$ Helmholtz Institute Mainz, Staudinger Weg 18, D-55099 Mainz, Germany\\
% ...
$^{28}$ Johannes Gutenberg University of Mainz, Johann-Joachim-Becher-Weg 45, D-55099 Mainz, Germany\\
% ...
$^{32}$ Liaoning Normal University, Dalian 116029, People's Republic of China\\
% ...
$^{49}$ State Key Laboratory of Particle Detection and Electronics, Beijing 100049, Hefei 230026, People's Republic of China\\
% ...
$^{57}$ University of Jinan, Jinan 250022, People's Republic of China\\
% ...
$^{63}$ University of Science and Technology of China, Hefei 230026, People's Republic of China\\
% ...
$^{66}$ University of Turin and INFN, (A)University of Turin, I-10125, Turin, Italy; (B)University of Eastern Piedmont, I-15121, Alessandria, Italy; (C)INFN, I-10125, Turin, Italy\\
$^{67}$ Uppsala University, Box 516, SE-75120 Uppsala, Sweden\\
% ...
$^{c}$ Also at the Novosibirsk State University, Novosibirsk, 630090, Russia\\
% ...
\vspace{0.4cm}
}\end{center}
\end{small}
}
\noaffiliation

\date{\today}

\begin{abstract}
  
The first amplitude analysis of the decay $D_s^+\to K^-K^+\pi^+\pi^0$ 
is presented using the data samples, corresponding to an integrated luminosity of 
6.32 fb$^{-1}$, collected with the BESIII detector at $e^+e^-$ center-of-mass energies 
between 4.178 and 4.226 GeV. More than 3000 events
selected with a purity of 97.5\% are used to perform the amplitude
analysis, and nine components are found necessary to describe the
data. Relative fractions and phases of the intermediate decays are
determined. With the detection efficiency estimated by the results
of the amplitude analysis, the branching fraction of
$D_s^+\to K^-K^+\pi^+\pi^0$ decay is measured to be $(5.42\pm0.10_{\rm 
stat.}\pm0.17_{\rm syst.})\%$.

\end{abstract}

\pacs{13.20.Fc, 12.38.Qk, 14.40.Lb}
\maketitle

\section{\boldmath Introduction}
Accurate measurement of $D_s^+$ decays are important for the studies
of other decay processes that are dominated by final states involving    
$D_s^+$ mesons, particularly for those of $B_s^0$ decays~\cite{lhcbbsds}.
The decay $D_s^+\to K^-K^+\pi^+\pi^0$ is a Cabibbo-favored decay 
(the inclusion of charge conjugated reactions is implied throughout this 
paper). Due to its large branching fraction (BF), it is usually selected 
as a ``tag mode" for the measurement of other decays of the $D_s^+$
meson~\cite{theref1,theref2,theref3,theref4,theref5,theref6}. 
However, the BF of the $D_s^+\to K^-K^+\pi^+\pi^0$ decay has a large systematic
uncertainty due to the poor knowledge of intermediate state
processes~\cite{2008abs,2013abs}. An amplitude analysis of this decay
is expected to provide a detailed understanding of its intermediate
structures and significantly improve the experimental precision of its
decay BF.

\begin{figure}[htbp]
\centering
    \mbox{
    \hskip -0.58cm
    \begin{overpic}[width=4.5cm,height=2.7cm,angle=0]{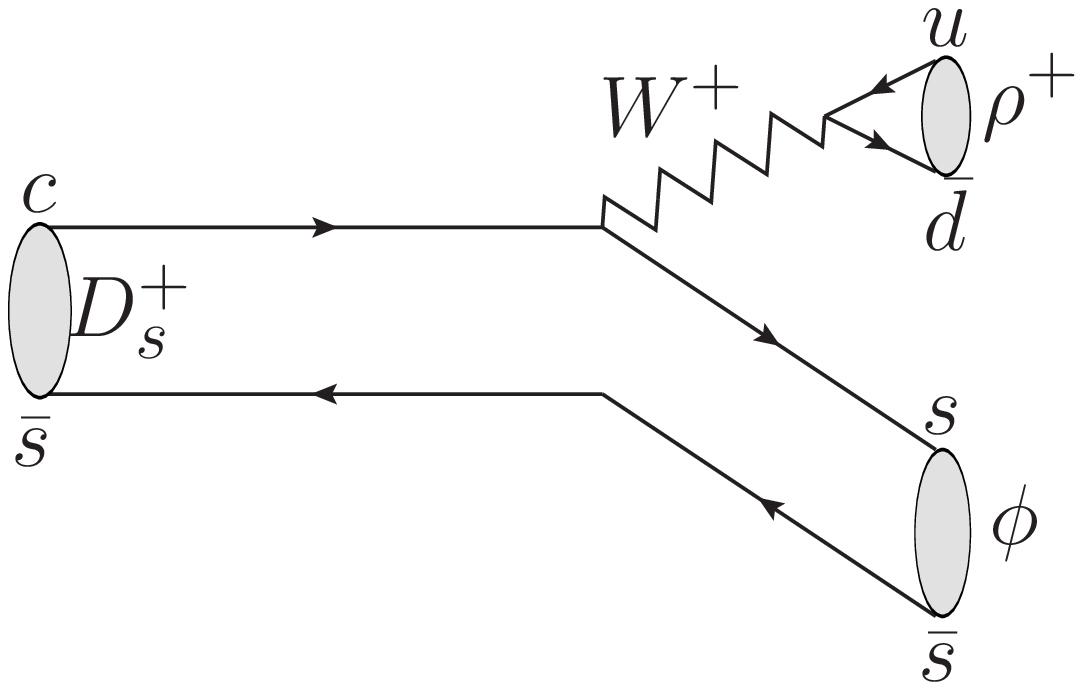}
    \put(40,-5){ (a) }
    \end{overpic}
    \hskip -0.27cm
    \begin{overpic}[width=4.5cm,height=2.7cm,angle=0]{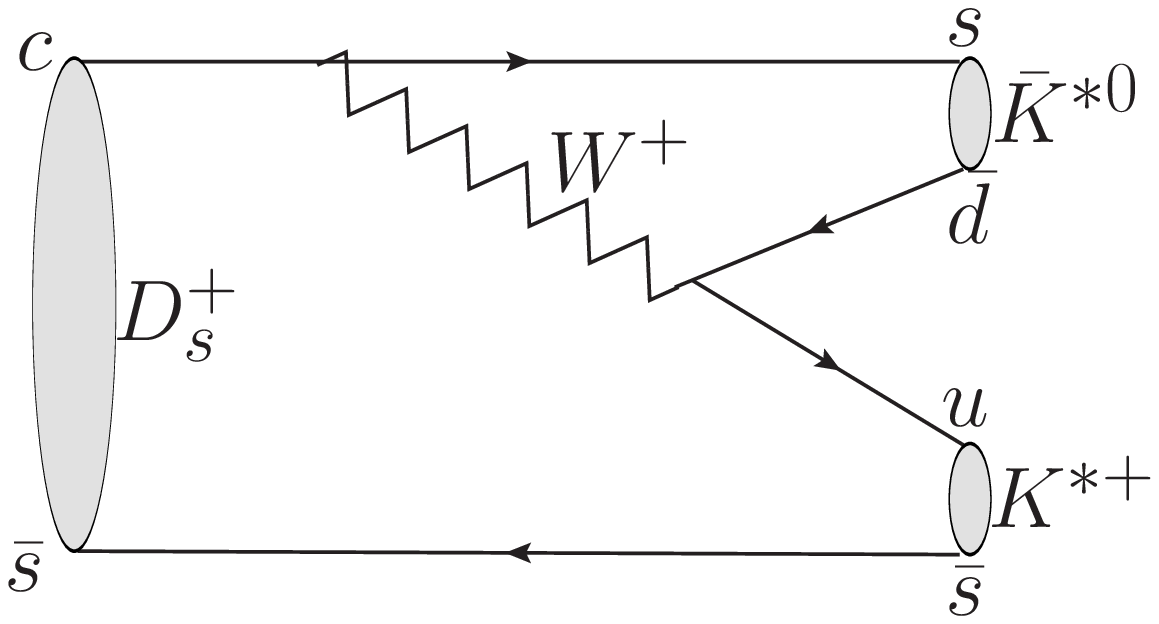}
    \put(40,-5){ (b) }
    \end{overpic}
    }
\caption{Decay diagrams of (a) $D_s^+\to \phi\rho^+$ and (b) $D_s^+\to \bar{K}^{*0}K^{*+}$ decays.}
\label{feynman}
\end{figure}

The four-body hadronic decays of $D_s^+$ mesons are dominated by
two-body intermediate processes, for example $D_s^+\to AP$ and 
$D_s^+\to VV$ decays, where $V$, $A$, and $P$ denote vector,
axial-vector, and pseudoscalar mesons, respectively. Measurements of
the BFs of these two-body decays are important to test theoretical
calculations~\cite{ceshijiu1,ceshijiu2,ceshijiu3,ceshijiu4} and to
better understand the decay mechanisms of the $D_s^+$ meson. 
In recent years, many measurements of $D_s^+\to PP$ and $D_s^+\to VP$ decays
have been reported~\cite{PDG}. However, there are few studies focusing 
on $D_s^+\to AP$ and $D_s^+\to VV$ decays. 
The amplitude analysis of $D_s^+\to AP$ decay allows the study 
of substructures involving $K_1(1270)$, $K_1(1400)$, and $f_1(1420)$ 
mesons. The measurements of the intermediate resonances $K_1(1270)$ 
and $K_1(1400)$ are also useful for understanding the mixing of these 
two axial-vector kaons~\cite{three}. 
For $D_s^+\to VV$, two processes, namely $D_s^+\to
\phi\rho^+$ and $D_s^+\to \bar{K}^{*0}K^{*+}$, which are represented
by the decay diagrams in Fig.~\ref{feynman}, can be studied in the
$D^+_s\to K^-K^+\pi^+\pi^0$ decay. The BF of the decay $D_s^+\to \phi\rho^+$
was measured to be $(8.4^{+1.9}_{-2.3})$\%~\cite{phirhopref} by
the CLEO experiment based on a data sample corresponding to an
integrated luminosity of 689 pb$^{-1}$ at the $\Upsilon(3S)$ and 
$\Upsilon(4S)$ resonances and at $e^+e^-$ center-of-mass energies 
($E_{\rm cm}$) just below and above the $\Upsilon(4S)$ resonance. 
The previous most precise determination of the BF of $D_s^+\to \bar{K}^{*0}K^{*+}$ decay,
$(7.2\pm2.6)$\%~\cite{kstkstref}, was performed by the ARGUS experiment
using a data sample of 432 pb$^{-1}$ collected at $E_{\rm cm} = 10.4$~GeV. 
The goal of the present analysis is to improve the precision of these measurements.

Moreover, a recent study~\cite{yufusheng} points out that the measured
values of the ratio of $K_1(1270)$ decay ($R_{K_1(1270)}= 
\frac{\mathcal{B}(K_1(1270)\to K^*\pi)}{\mathcal{B}(K_1(1270)\to
  K\rho)}$), which are listed in Table~\ref{etaao}, are inconsistent
between different experiments~\cite{jhepssss,2012D0kkpipi1,shiliu,shiliu0,shiliu1,shiliu2,2017D0kkpipi2}. 
They are expected to be identical under the narrow width approximation for
the $K_1(1270)$ meson and assuming $CP$ conservation in strong decays~\cite{yufusheng}. 
The decays related to this ratio may be observed in the $D_s^+\to K^-K^+\pi^+\pi^0$ decay.
\begin{table}[htb]
 \centering
 \caption{Values of $R_{K_1(1270)}$ determined by different
   experiments. Fit 1 and fit 2 refer to amplitude analyses performed
   with the mass and width of the $K_1(1270)^+$ meson fixed or
   left free in the fit, respectively.}
  \vspace{2mm} 
 \label{etaao}
\renewcommand\arraystretch{1.2}
 \begin{tabular}{cll}
\hline\hline
$R_{K_1(1270)}$         & Process                              & Experiment\\ \hline     
  $0.81\pm0.10$         & $D^0\to K^+K^-\pi^+\pi^-$            & LHCb~\cite{jhepssss} \\
  $1.18\pm0.43$         & $D^0\to K^-K_1^+(1270)$              & CLEO~\cite{2012D0kkpipi1} \\   
  $0.11\pm0.06$         & $D^0\to K^+K_1^-(1270)$              & CLEO~\cite{2012D0kkpipi1} \\   
  $0.19\pm0.10$         & $D^0\to K^-\pi^+\pi^+\pi^-$          & BESIII~\cite{shiliu} \\ 
  $0.24\pm0.04$         & $D^0\to K^-\pi^+\pi^+\pi^-$          & LHCb~\cite{shiliu0} \\  
  $0.45\pm0.05$         & $B^+\to J/\psi K^+\pi^+\pi^-$        & Belle~\cite{shiliu1} (Fit 1) \\ 
  $0.30\pm0.04$         & $B^+\to J/\psi K^+\pi^+\pi^-$        & Belle~\cite{shiliu1} (Fit 2) \\ 
  $0.38\pm0.13$         & $K^-p\to K^-\pi^-\pi^+p$             & ACCMOR~\cite{shiliu2} \\ 
  $0.45\pm0.14$         & $D^0\to K^-K_1^+(1270)$              & CLEO~\cite{2017D0kkpipi2} \\
\hline\hline
 \end{tabular}
\end{table}

In this paper, the first amplitude analysis of the decay $D_s^+\to K^-K^+\pi^+\pi^0$ 
is presented using data samples of 6.32 fb$^{-1}$ collected with the BESIII detector at center-of-mass 
energies between 4.178 and 4.226 GeV. 
The amplitude model is constructed with the covariant tensor
formalism~\cite{Zou2003} and described in Section~\ref{amplitude}.
The BF measurement is presented in Section~\ref{branchfractionsssss}.

\section{\boldmath Detector and data sets}
The BESIII detector is a magnetic spectrometer~\cite{BESIII,BESIII2}
located at the Beijing Electron Position Collider
(BEPCII)~\cite{BEPCII}. The cylindrical core of the BESIII detector
consists of a helium-based multilayer drift chamber (MDC), a plastic
scintillator time-of-flight system (TOF), and a CsI(Tl)
electromagnetic calorimeter (EMC), which are all enclosed in a
superconducting solenoidal magnet providing a 1.0 T magnetic
field. The solenoid is supported by an octagonal flux-return yoke with
resistive-plate counter muon identifier modules interleaved with
steel. The acceptance of charged particles and photons is 93\% over
the 4$\pi$ solid angle. The resolution of charged-particle momentum at
1 GeV/$c$ is 0.5\% while that of the specific ionization energy loss
($dE/dx$) is 6\% for electrons from Bhabha scattering. The EMC
measures photon energies with a resolution of 2.5\% (5\%) at 1 GeV in
the barrel (end-cap) region. The time resolution of the TOF barrel
part is 68 ps. The end-cap TOF system was upgraded in 2015 with
multi-gap resistive plate chamber technology, providing a time
resolution of 60 ps~\cite{shisiya,shiwuya}.

The data samples used in this analysis contain a total integrated 
luminosity of 6.32~fb$^{-1}$ collected at center-of-mass energies 
between $E_{\rm cm}=4.178$ and 4.226~GeV with the BESIII detector. 
The integrated luminosity of each data sample is shown in
Table~\ref{mrec}. In this energy region, pairs of $D_{s}^{\pm}
D^{*\mp}_{s}$ mesons are produced. The $D^{*\pm}_{s}$ meson
predominantly decays to $D_{s}^{\pm}\gamma$ (93.5\%), and only a small
fraction decays to $D_{s}\pi^{0}$ (5.8\%)~\cite{PDG}. A double-tag
(DT) technique is employed to measure the absolute BF of the $D_s^+$
decays~\cite{ddbbttkk}. First, the $D_s^-$ meson is fully
reconstructed in one of the following decay modes: $D_s^-\to K_S^0
K^-$, $D_s^-\to K^+K^-\pi^-$, $D_s^-\to K_S^0K^-\pi^0$, $D_s^-\to
K_S^0K^-\pi^+\pi^-$, $D_s^-\to K_S^0K^+\pi^-\pi^-$, $D_s^-\to
\pi^-\eta_{\gamma\gamma}$, $D_s^-\to
\pi^-\eta^{\prime}_{\pi^+\pi^-\eta_{\gamma\gamma}}$, and $D_s^-\to
K^-\pi^+\pi^-$. These are referred to as single-tag (ST)
events. Second, the $D_s^+\to K^-K^+\pi^+\pi^0$ decay events are
selected. 

Generic Monte Carlo (MC) simulated event samples are produced with the
GEANT4-based~\cite{geant4,geant4pak} software at $E_{\rm
  cm}=4.178-4.226$ GeV. The samples include all known open charm
decays; the continuum process ($e^+e^-\to q\bar{q}$, $q=u$, $d$, and
$s$); Bhabha scattering; the $\mu^+\mu^-$, $\tau^+\tau^-$, and
diphoton processes; and the $c\bar{c}$ resonances ($J/\psi$,
$\psi(3686)$, and $\psi(3770)$) via initial-state radiation (ISR). The
open charm processes are generated using {\sc conexc}~\cite{conexc},
and their subsequent decays are modeled by {\sc
  evtgen}~\cite{evtgennn} with known BFs from the Particle Data Group
(PDG)~\cite{PDG}. The simulation of ISR production of $J/\psi$,
$\psi(3686)$, and $\psi(3770)$ mesons is performed with {\sc
  kkmc}~\cite{kkmckaka}. The effects of final-state radiation (FSR)
from charged tracks are simulated by {\sc photos}~\cite{photosss}. The
remaining unknown decays are generated with the {\sc lundcharm}
model~\cite{lundcharm}. The generic MC sample is used to study backgrounds
and determine the efficiencies of tag modes and the signal mode for
the BF measurement, in which our amplitude analysis model is 
used to generate the signal mode events.

A phase-space (PHSP) MC sample is produced with the
$D_s^+$ meson decaying to $K^-K^+\pi^+\pi^0$ generated with a uniform
distribution and the $D_s^-$ meson decaying to the tag modes.  Initially,
the PHSP MC sample is used to calculate the normalization integral
used in the determination of the amplitude model parameters in the fit to
data. Then, the signal MC sample is re-generated with the $D_s^+$ meson decaying to $K^-K^+\pi^+\pi^0$ 
using the amplitude model and the $D_s^-$ meson decaying to the tag modes.
The normalization integral performed with signal MC samples results 
in more accurate fit parameters of magnitudes and phases and improves 
the computational efficiency of the MC integration. The signal MC sample 
is also used to calculate the goodness of the fit in this analysis. 
The PHSP MC sample is used to determine the efficiency mentioned in 
Section~\ref{buzhegenage}.

\section{\boldmath Event selection}
Charged tracks except for those from $\kshort$ decays are required to
have a distance of closest approach to the interaction point (IP)
within 1~cm in the transverse plane and within 10~cm along the 
MDC axis ($z$ axis). The polar angle of the charged track with respect 
to the $z$ axis $\theta$ is required to satisfy $|\cos \theta| < 0.93$.
Kaons and pions are identified by combining the $dE/dx$ information in 
the MDC and the time-of-flight from the TOF. If the probability of the 
kaon hypothesis is larger than that of the pion hypothesis, the track is 
identified as a kaon. Otherwise, the track is identified as a pion. Particle
identification (PID) is not performed for the $\pi^+$ or $\pi^-$ from
$\kshort$ decays.

The $\pi^0$ and $\eta$ candidates are reconstructed via diphoton
decays. The timing of the electromagnetic showers in the EMC is required
to be within [0,700] ns of the trigger time, and the deposited energy
must be greater than 25 (50) MeV in the barrel (endcap) region of the
EMC. Good showers must satisfy $|\cos \theta| < 0.80$ ($0.86 < |\cos \theta < 0.92$) in
the barrel (endcap) and be more than $20^{\circ}$ from the nearest charged
track. The unconstrained invariant masses of $\pi^0$, $\eta$ and
$\eta^{\prime}$ ($\eta^{\prime}\to\pi^+\pi^-\eta_{\gamma\gamma}$) are
required to be within [115,~150] MeV/$c^{2}$, [500,~570]~MeV/$c^{2}$,
and [946,~970] MeV/$c^{2}$, respectively. A kinematic fit is performed
to constrain $M_{\gamma\gamma}$ to the known $\pi^0$ ($\eta$) mass, and
the $\chi^{2}$ of the corresponding fit is required to be less than 30 (20)
for $\pi^0$ ($\eta$) candidates.

The $\kshort$ candidates are reconstructed in the decay
$\kshort\to\pi^+\pi^-$. Two oppositely charged tracks with distances
of closest approach to the IP less than 20~cm along the $z$ axis are
assigned as $\pi^+\pi^-$ without further PID requirements. A
constrained vertex fit of each pair of tracks is performed. Candidate $K_S^0$ particles are required to have the $\chi^2$ of the vertex fit less than 100 and
an invariant mass of the $\pi^+\pi^-$ pair ($M_{\pi^+\pi^-}$)
in the range [487,~511] MeV/$c^{2}$. In
the case of the decay modes $D_s^-\to K_S^0 K^-\pi^0$, $D_s^-\to
K_S^0K^-\pi^+\pi^-$ and $D_s^-\to K_S^0K^+\pi^+\pi^-$, the decay
length of the $K_S^0$ candidates obtained with the secondary vertex
fit~\cite{sec-vtx} must be at least two times its fit uncertainty. For the
$D_s^-\to K^-\pi^+\pi^-$ process, 
$M_{\pi^+\pi^-}$ is required to be
outside of the range [487,511] MeV/$c^{2}$, to remove possible misidentified events of $D_s^-\to K_S^0K^-$.

To identify the process $e^{+}e^{-}\rightarrow D^{*-}_{s}D_{s}^{+}$,
 the recoil mass $M_{\rm rec}$ of $D_s^-$ candidates is defined as
\begin{equation}\label{eawoc}
M_{\rm rec}=\sqrt{\Big (E_{\rm cm}- \sqrt{|\vec{p}_{D_s^-}|^2 + m^2_{D_s^-}}\Big )^2 -|\vec{p}_{D_s^-}|^2 }\,,
\end{equation}
where $m_{D_s^-}$ is the nominal $D_s^-$ mass~\cite{PDG} and
$\vec{p}_{D_s^-}$ is the momentum of the $D_s^-$ candidate. The values
of $M_{\rm rec}$ are required to be in the
regions depending on the center-of-mass energy as listed in Table~\ref{mrec}.
The $D_s^-$ mass windows for the
eight tag modes are shown in Table~\ref{tagwindow}.
\begin{table}[htp]
\begin{center}
  \caption{The integrated luminosities ($\mathcal{L}_\text{int}$) and the
    requirements on $M_{\rm rec}$ for various energies. $M_{\rm rec}$
    is defined in Eq.~\ref{eawoc}.}
  \vspace{2mm} 
\renewcommand\arraystretch{1.2}
\begin{tabular}{ccc}
\hline\hline
 $E_{\rm cm}$ (GeV) &  $\mathcal{L}_\text{int}$ (pb$^{-1}$) & $M_{\rm rec}$ (GeV/$c^2$)\\\hline 
   4.178            &  $3189.0$   & [2.050,~2.180]\\ 
   4.189            &  $526.7$    & [2.048,~2.190]\\ 
   4.199            &  $526.0$    & [2.046,~2.200]\\ 
   4.209            &  $517.1$    & [2.044,~2.210]\\ 
   4.219            &  $514.6$    & [2.042,~2.220]\\ 
   4.226            &  $1047.3$   & [2.040,~2.220]\\
\hline\hline
\end{tabular} 
\label{mrec}  
\end{center}  
\end{table}   

\begin{table}[htp]
  \centering
  \caption{The $D_s^-$ mass requirements for the eight tag modes.}
  \vspace{2mm} 
  \label{tagwindow}
\renewcommand\arraystretch{1.2}
  \begin{tabular}{lc}
\hline\hline
  Tag mode                           & Mass window (GeV/$c^2$)\\ 
  \hline
  $D_s^-\to K_S^0 K^-$                                     &  [1.948,~1.991]         \\ 
  $D_s^-\to K^+K^-\pi^-$                                   &  [1.950,~1.986]         \\ 
  $D_s^-\to K_S^0K^-\pi^0$                                 &  [1.946,~1.987]         \\ 
  $D_s^-\to K_S^0K^-\pi^+\pi^-$                            &  [1.958,~1.980]         \\ 
  $D_s^-\to K_S^0K^+\pi^-\pi^-$                            &  [1.953,~1.983]         \\ 
  $D_s^-\to \pi^-\eta_{\gamma\gamma}$                      &  [1.930,~2.000]         \\ 
  $D_s^-\to \pi^-\eta^{\prime}_{\pi^+\pi^-\eta_{\gamma\gamma}}$ &  [1.940,~1.996]         \\ 
  $D_s^-\to K^-\pi^+\pi^-$                                 &  [1.953,~1.983]         \\ 
\hline\hline
  \end{tabular}
\end{table}

The $D_s^+$ meson decays with invariant masses $M_{D_s^+}$ in the region
[1.87,~2.06] GeV/$c^2$ are selected. Good vertex fits of all charged
tracks on both the signal and the tag side are required. A
multi-constraint kinematic fit of $e^+e^-\to D_s^{*\pm}D_s^{\mp}\to
\gamma D_s^{\pm}D_s^{\mp}$ with $D_s^-$ decaying to one of the tag modes
and $D_s^+$ decaying to the signal mode is performed. The set of constraints
including four-momentum conservation in the $e^+e^-$ system and the
mass constraints of the $\pi^0$ meson, the $D_s^+$ meson, the $D_s^-$ meson
and the $D_s^{*\pm}$ meson is labeled $C_1$. Based on the requirements of $C_1$,
a set of constraints $C_2$ is defined by excluding the signal $M_{D_s^+}$
constraint, and $C_3$ is defined by excluding the mass constraints of the
$D_s^{\pm}$ meson on both the signal and tag sides.

If there are multiple candidate combinations in an event, the
candidate with the minimum $\chi^2$ of the $C_2$ kinematic fit
($\chi^2_{C_2}$) is chosen. A good $C_1$ kinematic fit is required. To
reduce the background while avoiding peaking background which is caused by
constraining the mass of the $D_s^{\pm}$ meson ($M_{D_s^{\pm}}$), 
the $\chi^2$ of the $C_3$ kinematic fit ($\chi^2_{C_3}$) is required to be less
than 250.

The classes of background events, which are listed in
Table~\ref{mis1}, are rejected. For backgrounds categorized as (a), (b) and (c), 
a $\pi^0$ from the $D_s^-$ decay is wrongly associated to the $D_s^+$ meson on the opposite side.
These are vetoed if the $\chi^2$ of the $C_1$ kinematic fit ($\chi^2_{C_1}$)
of the reconfigured combination is better than that of the
original. For backgrounds categorized as (d), the events with $D^+\to
K^-\pi^+\pi^+$ decay versus $D^-\to K^+K^-\pi^-\pi^0$ decay are
wrongly reconstructed as $D_s^-\to K^-\pi^+\pi^-$ decay versus
$D_s^+\to K^+K^-\pi^+\pi^0$ decay, when a $\pi^-$ meson from $D^-$
decay is exchanged with a $\pi^+$ meson from $D^+$ decay. If the
reconstructed $D^{\pm}$ masses of the signal and the tag modes fall in
the region within 0.055 GeV/$c^2$ of the nominal $M_{D^{\pm}}$, the
events are rejected. For background categories (e) and (f), events
with $K_S^0K^+K^-$ satisfying $|M_{K_S^0K^+K^-}-M_{D^0}^{\rm
  PDG}|<0.045$~GeV/$c^2$ are rejected, where $M_{D^0}^{\rm PDG}$ is
the nominal mass of $D^0$~\cite{PDG}. For background category (g), the
wrong signal combination survives due to exchanging the $\pi^0$ meson
from $D^0$ decay and the $\pi^-$ meson from $\bar{D}^0$ decay and
misidentifying the $\pi^-$ meson as a $K^-$ meson. They are suppressed by
rejecting events satisfying $|M_{K^-\pi^+\pi^0}-M_{D^0}^{\rm PDG}|<$ 0.055~GeV/$c^2$ and
$|M_{K^+\pi^-\pi^+\pi^-}-M_{D^0}^{\rm PDG}|<0.055$ 
GeV/$c^2$. Background type (h) events are suppressed by applying a
veto on events with $|M_{K^-\pi^+\pi^0}-M_{D^0}^{\rm PDG}|<0.045$
GeV/$c^2$.
\begin{table}[htp]
  \centering
  \caption{Misreconstructed background processes.}
  \vspace{2mm} 
  \label{mis1}
\renewcommand\arraystretch{1.2}
  \begin{tabular}{cll}
\hline\hline
Category &  \multicolumn{2}{c}{Background} \\
  \hline
(a)  &   $D_s^+\to K^+K^-\pi^+$, & $D_s^-\to \pi^-\pi^0\eta$ \\
(b)  &   $D_s^+\to K^+K^-\pi^+$, & $D_s^-\to \pi^-\pi^0\eta^{\prime}$ \\
(c)  &   $D_s^+\to K^+K^-\pi^+$, & $D_s^-\to K^-\pi^-\pi^+\pi^0$ \\
(d)  &   $D^+\to K^-\pi^+\pi^+$, & $D^-\to K^+K^-\pi^-\pi^0$ \\
(e)  &   $\bar{D}^0\to K_S^0K^+K^-$, & $D^0\to K^-\pi^+\pi^0$ \\
(f)  &   $\bar{D}^0\to K_S^0K^+K^-$, & $D^0\to K^-\pi^+\pi^0\pi^0$ \\
(g)  &   $D^0\to K^-\pi^+\pi^0$, & $\bar{D}^0\to K^+\pi^-\pi^+\pi^-$ \\
(h)  &   $\bar{D}^0\to K^+\pi^-\pi^0$, & $D^0\to K^-\pi^+\pi^0$ \\
\hline\hline
  \end{tabular}
\end{table}

Events containing a possible mis-formed $\pi^0$ meson on the signal
side are also rejected.  Events in which the invariant mass of the higher-energy
photon from the signal side combined with a photon from the 
$D_s^* \rightarrow D_s \gamma$ decay is within [0.12,~0.15] GeV/$c^2$ and with $|dM_{\rm
  recombined}|<|dM|$ are rejected, where $dM$ is the mass difference
between the signal $D_s^+$ meson and the tagged $D_s^-$ meson, and
$dM_{\rm recombined}$ is the corresponding mass difference with the signal $\pi^0$ replaced by
the recombined $\pi^0$. A veto is also applied to reject events with recombined mass of the higher-energy photon from the signal side and the photon from the tag side falling within [0.12,~0.15]~GeV/$c^2$.

\begin{figure*}[htbp]
\centering
    \hskip -0.4cm \mbox{
    \begin{overpic}[width=5.7cm,height=4.6cm,angle=0]{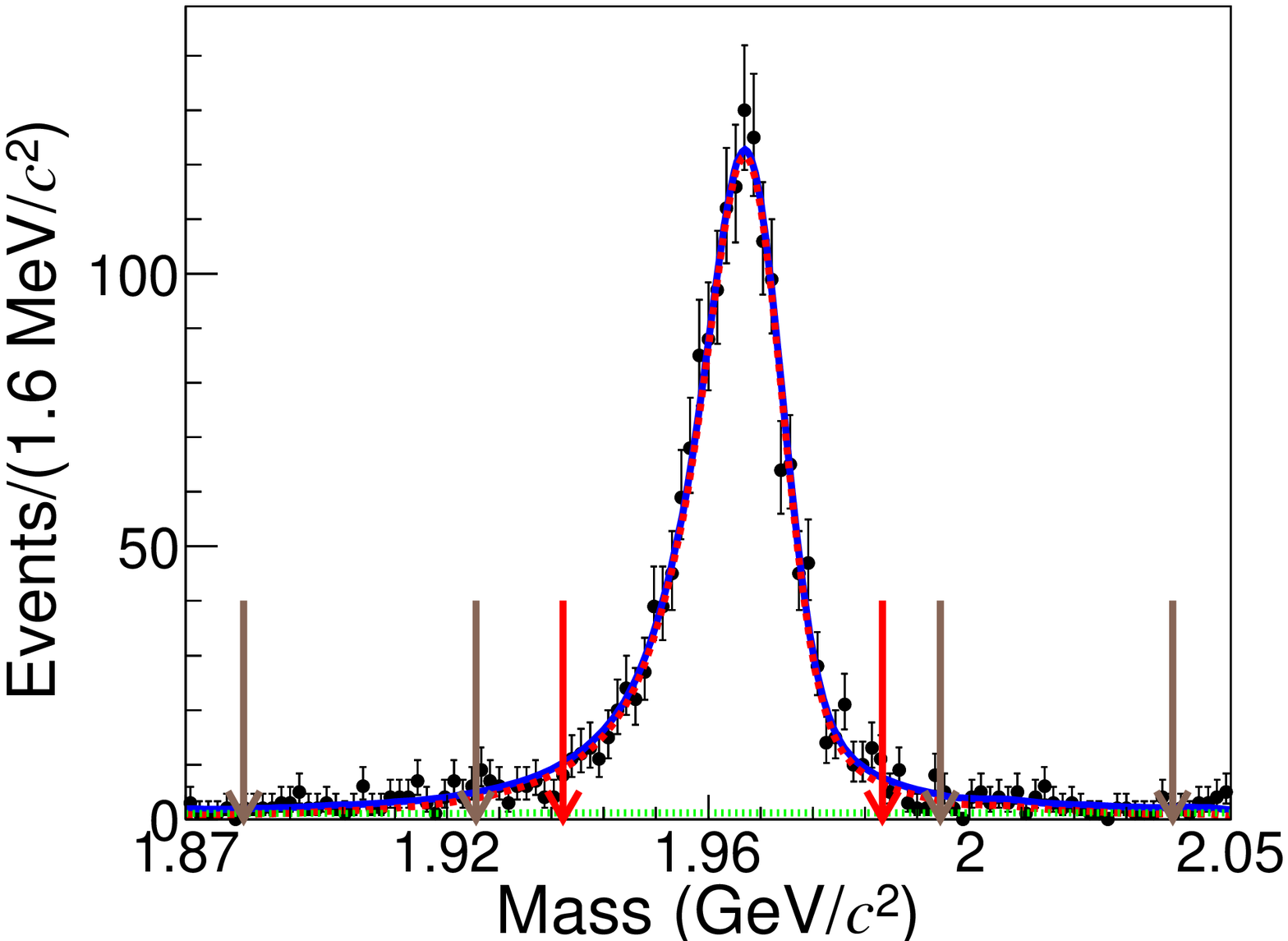}
    \put(65,60){{\large (a) }}
    \end{overpic}
    \begin{overpic}[width=5.7cm,height=4.6cm,angle=0]{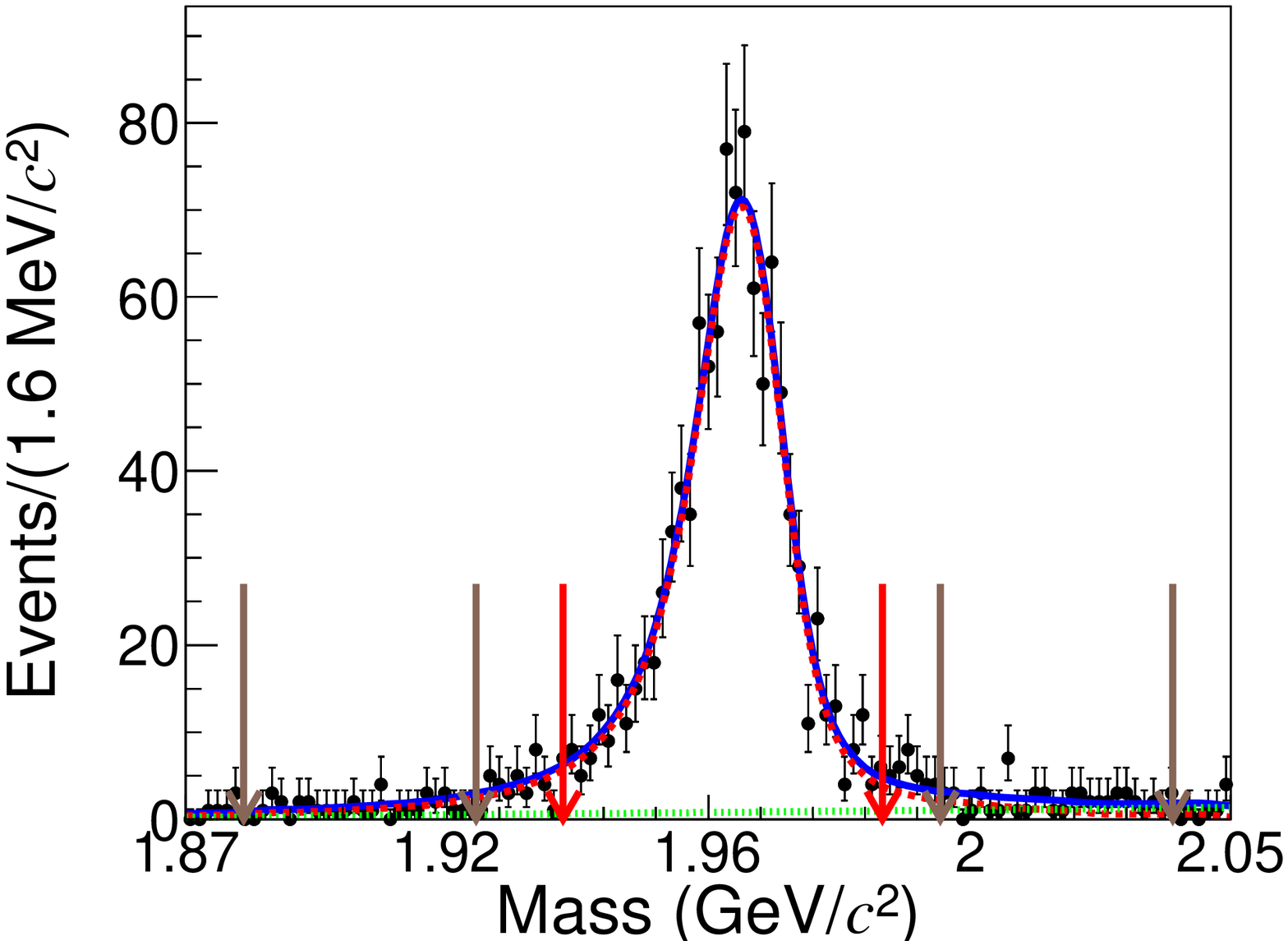}
    \put(65,60){{\large (b) }}
    \end{overpic}
    \begin{overpic}[width=5.7cm,height=4.6cm,angle=0]{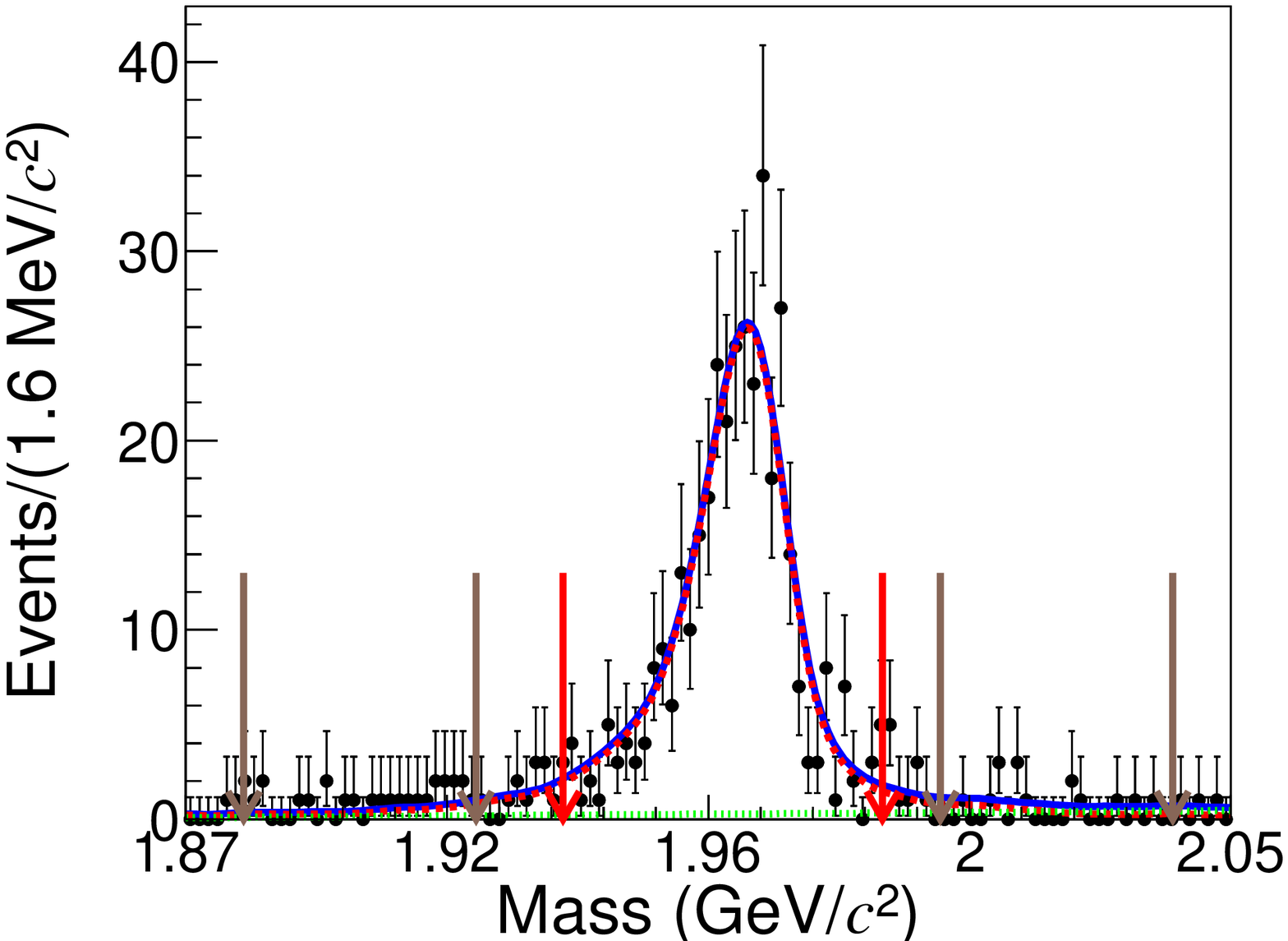}
    \put(65,60){{\large (c) }}
    \end{overpic}    }
\caption{Fits to the invariant mass spectra of the signal $D_s^+$
  candidates for data samples collected at center-of-mass energies (a)
  4.178 GeV, (b) 4.189-4.219 GeV and (c) 4.226 GeV. The black dots
  with error bars represent data. The red dotted line represents the
  MC-simulated shape convolved with a Gaussian function. The green
  dashed lines are the fitted backgrounds. The blue solid line
  represents the total fitted shape. The red arrows represent the
  requirements applied in the amplitude analysis and the brown arrows
  represent the sideband region.} 
\label{fitfig333}
\end{figure*}

After the full selection, the invariant mass spectra of the signal $D_s^+$
candidates for data samples collected at center-of-mass energies
4.178-4.226 GeV are shown in Fig.~\ref{fitfig333}, together with fits to the mass spectra. 
There are 1708, 1024, and 356 events retained in the signal region [1.935,~1.99] GeV/$c^2$ for the amplitude analysis with purities, $w_{\rm sig}$, of $97.7\pm0.4$\%, $97.3\pm0.5$\% and $97.5\pm0.8$\% at $E_{\rm cm}=$ 4.178 GeV, 4.189-4.219 GeV, and 4.226 GeV, respectively.
Studies of the generic MC samples show that peaking
background is negligible. 
The background description by the generic MC has been verified by comparisons 
of data with the generic MC samples in the sideband regions 
[1.88,~1.92] GeV/$c^2$ and [2.00,~2.04] GeV/$c^2$.
A good agreement is found, and the generic MC samples are used to model the
residual background contamination in the signal region.
The four-momenta of the final state particles after a two-constraint kinematic fit to the
signal candidate, constraining the $D_s^+$ mass and $\pi^0$ mass
to their known values~\cite{PDG}, are used to perform the amplitude
analysis.

\section{\boldmath Amplitude analysis}\label{amplitude}
The amplitude analysis of $D_s^+\to K^-K^+\pi^+\pi^0$ decay is
performed by using an unbinned maximum likelihood fit. The likelihood
function is constructed with the probability density function (PDF) described in the following,
in which the momenta of the four daughter particles are used as
inputs. 

\subsection{\boldmath Likelihood Function
Construction}\label{buzhegenage} The PDF used to construct the
likelihood of the amplitude is given by \begin{eqnarray}
\begin{aligned} \label{phspinter} f_S(p_j) = \frac{\epsilon(p_j)\vert
M(p_j)\vert^2R_4(p_j)}{\int\epsilon(p_j)\vert
M(p_j)\vert^2R_4(p_j)dp_j}\,, \end{aligned} \end{eqnarray} where $p_j$
is the set of the final state particles' four momenta, and
$\epsilon(p_j)$ is the detection efficiency parameterized in terms of
the final state particles' four momenta. The PDF $f_S(p_j)$ is
normalized by the integration. The standard element of the four-body
PHSP~\cite{Zou2003} is defined as \begin{eqnarray} \begin{aligned}
R_4(p_j)dp_j=\delta^4\Big (p_{D_s^+}-\sum^4_{j=1} p_j\Big
)\prod^4_{j=1}\frac{d^3p_j}{(2\pi)^3 2E_j}\,, \end{aligned}
\end{eqnarray} where $j$ runs over the four daughter particles and $E_j$
is the energy of particle $j$.

This analysis uses an isobar model formulation, where
the signal decay amplitude, $M(p_j)$, is represented as a coherent sum of many two-body amplitude modes
\begin{eqnarray}
\begin{aligned} 
\label{eq:total_amplitude}
M(p_j) = \sum_n c_nA_n(p_j)\,,            
\end{aligned}
\end{eqnarray}
where $c_n$ is written in the polar form as $\rho_ne^{i\phi_n}$ ($\rho_n$ and $\phi_n$ are the magnitude  
and phase for the $n^{\text{th}}$ amplitude, respectively). $A_n(p_j)$ is the $n^{\text{th}}$ amplitude function modeled as
\begin{align}
A_n(p_j)&=P^1_n(m_1)P^2_n(m_2)S_n(p_j)X^1_n(p_j)X^2_n(p_j)X^{D_s^+}_n(p_j)\,,
\end{align}           
where the indices 1 and 2 correspond to the two intermediate
resonances, respectively. $X^{D_s^+}_n(p_j)$ is the Blatt-Weisskopf
barrier factor~\cite{Zou2003,shilei1,shilei2,shijiu} for the $D_s^+$ meson,
while $P^{1,2}_n(m_1,m_2)$ and $X^{1,2}_n(p_j)$ are the propagators
and Blatt-Weisskopf barrier factors of the intermediate resonances 1
and 2, respectively. For non-resonant states, the propagator is set to
unity, which can be regarded as a very broad resonance. 
$S_n(p_j)$ is the spin factor which is constructed with the covariant
tensor formalism~\cite{Zou2003}.

The 2.5\% background contribution is described by the background PDF:
\begin{eqnarray}
\begin{aligned}\label{bkgpdf} f_{B}(p_j) =
\frac{B(p_j)R_4(p_j)}{\int B(p_j)R_4(p_j)dp_j}\,.
\end{aligned}
\end{eqnarray}

The background events in the signal region from the generic MC sample are used to model
the corresponding background in data. 
The background shape $B(p_j)$ is derived using a multi-dimensional kernel density estimator~\cite{25af} with five independent Lorentz invariant variables ($M_{K^-K^+}$, $M_{\pi^+\pi^0}$, $M_{K^-\pi^0}$, $M_{K^-\pi^+}$, and $M_{K^-K^+\pi^0}$) implemented in RooFit~\cite{24fdf}, which models the distribution
of an input dataset as a superposition of Gaussian kernels. This background PDF is then
added to the signal PDF incoherently, and the combined PDF is written as
\begin{flalign} \qquad \,f_T(p_j)&=w_{\rm
sig}f_S(p_j)+(1-w_{\rm sig})f_{B}(p_j)& \nonumber\\ \qquad
\,&=w_{\rm sig}\frac{\epsilon(p_j)\vert
M(p_j)\vert^2R_4(p_j)}{\int\epsilon(p_j)\vert
M(p_j)\vert^2R_4(p_j)dp_j}& \nonumber\\ \qquad \,&\quad+(1-w_{\rm
sig})\frac{B(p_j)R_4(p_j)}{\int B(p_j)R_4(p_j)dp_j}\,,&
\end{flalign}
where the factor $\epsilon(p_j)$ in the numerator can be
taken out as in Eq.~\ref{combine2}. In this way, the $\epsilon(p_j)$
term, which is independent of the fitted variables, is a constant and
can be dropped in the likelihood fit. For the determination of
$\epsilon(p_j)$, totally 300 million PHSP MC events at $E_{\rm cm}=$ 4.178
GeV, 4.189-4.219 GeV and 4.226 GeV are generated, and near 15 million 
events are selected with the event selection. The background shape is
determined from the selected generic MC events; hence one has to divide the
background function by the efficiency,
$B_{\epsilon}\equiv B/\epsilon$.  The value
$\epsilon(p_j)$ is calculated as the fraction of selected PHSP MC events
in the five-dimensional space ($M_{K^-K^+}, M_{\pi^+\pi^0}, M_{K^-\pi^0},
M_{K^-\pi^+}, M_{K^-K^+\pi^0}$) with
$10\times10\times10\times10\times10$ bins.
 
The combined PDF becomes \begin{flalign}\label{combine2}
&f_T(p_j)=\epsilon(p_j)R_4(p_j)\Bigg[w_{\rm sig}\frac{\vert
M(p_j)\vert^2}{\int\epsilon(p_j)\vert M(p_j)\vert^2R_4(p_j)dp_j}&
\nonumber\\ &\qquad \qquad+ (1-w_{\rm
sig})\frac{B_{\epsilon}(p_j)}{\int\epsilon(p_j)B_{\epsilon}(p_j)R_4(p_j)dp_j}\Bigg]\,.&
\end{flalign}

The corresponding likelihood function is defined as 
\begin{eqnarray}  
\begin{aligned}\label{weihe}
L_i=\prod_{k_{i}=1}^{N_{\rm data}^i} f_T^{k_i}(p_j)\,,
\end{aligned}          
\end{eqnarray}
where $i$ denotes the data sample, $k_i$ runs over each event and
$N_{\rm data}^i$ is the number of events in data sample $i$. The
log-likelihood is used to perform the max-likelihood calculation.

The PDFs and the efficiencies are considered separately for three data samples with $E_{\rm cm}=$ 4.178 GeV, 4.189-4.219 GeV, and 4.226 GeV, corresponding to how the data were collected. 
Therefore, the
log-likelihood functions for the three samples are summed up
\begin{eqnarray}
\begin{aligned} 
\ln\mathcal{L}=\sum_{i=1}^{N=3}\ln L_i\,.
\end{aligned}
\end{eqnarray}

The normalization integrals in the denominator of Eq.~\ref{combine2} are obtained by summing over an MC sample 
\begin{align}\label{eq:MCintegration_signal} 
\int\epsilon(p_j)\vert M(p_j)\vert^2R_4(p_j)dp_j\approx \frac{1}{N_{\text{MC}}}\sum^{N_{\text{MC}}}_{k=1} \frac{\vert M(p_j^k)\vert^2}{\vert M^{\text{gen}}(p_j^k)\vert^2}\,,
\end{align}           
\begin{align}\label{eq:MCintegration_bkg}
\int\epsilon(p_j) B_{\epsilon}(p_j)R_4(p_j)dp_j\approx \frac{1}{N_{\text{MC}}}\sum^{N_{\text{MC}}}_{k=1} \frac{B_{\epsilon}(p_j^k)}{\vert M^{\text{gen}}(p_j^k)\vert^2}\,,
\end{align}           
where $N_{\text{MC}}$ is the number of the selected MC events and
$M^{\text{gen}}(p_j)$ is the amplitude that is set with the parameters
used to generate the signal MC sample, which are initially obtained from the results
using the PHSP MC integration. $M^{\text{gen}}(p_j)$ is a constant
over the whole PHSP. Then with the results obtained from the fit to
data, the signal MC sample is generated and used in MC integration.
Totally 12 million PHSP MC events and 10 million signal MC events are selected at $E_{\rm cm}=$ 4.178 GeV, 4.189-4.219 GeV and 4.226 GeV with satisfying all selection criteria as those of the data sample. 

The effect from the PID, tracking and reconstruction
efficiency differences between data and simulation is considered by multiplying the
weight of the MC event by a factor $\gamma_{\epsilon}$, which is
calculated as
\begin{eqnarray}  
\begin{aligned}\label{wochache}
\gamma_{\epsilon}(p_j)=\prod_{j}\frac{\epsilon_{j,\rm data}(p_j)}{\epsilon_{j,\text{MC}}(p_j)}\,, 
\end{aligned}  
\end{eqnarray}
where $j=K^{\mp}$, $K^{\pm}$, $\pi^{\pm}$ and $\pi^0$. The signal MC integration becomes
\begin{align}
\int\epsilon(p_j)\vert M(p_j)\vert^2R_4(p_j)dp_j\approx \frac{1}{N_{\text{MC}}}\sum^{N_{\text{MC}}}_{k=1} \frac{\vert M(p_j^k)\vert^2\gamma_{\epsilon}(p_j^k)}{\vert M^{\text{gen}}(p_j^k)\vert^2}\,.
\end{align}
\subsubsection{Spin Factors}
For a decay process of the form
$a \rightarrow bc$, $p_a$, $p_b$, $p_c$ are used 
to denote the momenta of the particles $a$, $b$, $c$, respectively. 
The spin projection operator $P^{(S)}_{\mu_1...\mu_{S}\nu_1...\nu_{S}}(a)$, for a resonance $a$ with spin $S = 0,1,2$ and four-momentum $p_a$ is given by
\begin{align} 
\begin{aligned}
P^{(0)}(a) &= 1\,,\\ 
P^{(1)}_{\mu\mu^{\prime}}(a) &= -g_{\mu\mu^{\prime}}+\frac{p_{a,\mu}p_{a,\mu^{\prime}}}{p^2_a}\,,\\ 
P^{(2)}_{\mu\nu\mu^{\prime}\nu^{\prime}}(a) &= \frac{1}{2}\Big (P^{(1)}_{\mu\mu^{\prime}}(a)P^{(1)}_{\nu\nu^{\prime}}(a)+P^{(1)}_{\mu\nu^{\prime}}(a)P^{(1)}_{\nu\mu^{\prime}}(a)\Big )\\
&\quad -\frac{1}{3}P^{(1)}_{\mu\nu}(a)P^{(1)}_{\mu^{\prime}\nu^{\prime}}(a)\,,
\end{aligned}
\end{align}
where $g_{\mu\mu^{\prime}}$ is the Minkowski metric.

The covariant tensors $\tilde{t}^{(L)}_{\mu_1...\mu_l}(a)$~\cite{Zou2003} for the final states of pure orbital angular momentum $L$ are constructed from the relevant momenta $p_a, p_b, p_c$:
\begin{align} 
\begin{aligned}
\tilde{t}^{(L)}_{\mu_1...\mu_l}(a) &= (-1)^{L}P^{(L)}_{\mu_1...\mu_{L}\nu_{1}...\nu_{L}}(a)r_a^{\nu_{1}}\cdots r_a^{\nu_{L}}\,, 
\end{aligned}
\end{align}
where $r_a\,=\,p_b - p_c$. The corresponding covariant tensors with $L=0,1,2$ are given as 
\begin{align} 
\begin{aligned}
\tilde{t}^{(0)}(a)&= 1\,,\\ 
\tilde{t}^{(1)}_{\mu}(a)&= -P^{(1)}_{\mu\mu^{\prime}}(a)r_a^{\mu^{\prime}}\,,\\  
\tilde{t}^{(2)}_{\mu\nu}(a)&= P^{(2)}_{\mu\nu\mu^{\prime}\nu^{\prime}}(a)r_a^{\mu^{\prime}}r_a^{\nu^{\prime}}\,. 
\end{aligned}
\end{align}

The eleven types of decay modes used in this analysis are listed in Table~\ref{table:spin_factors}. 
\begin{table*}[hbtp]
 \caption{Spin factor for each decay chain. All operators, i.e.~$\tilde{t}$, have the same definitions as Ref.~\cite{Zou2003}. 
   Scalar, pseudo-scalar, vector, and axial-vector  states are denoted 
   by $S$, $P$, $V$, and $A$, respectively. $[S]$, $[P]$, and $[D]$
   indicate the orbital angular momenta $L$= 0, 1, and 2 of the
   two-body final states, respectively.}
\label{table:spin_factors}
 \tabcolsep 7pt
 \renewcommand\arraystretch{1.2}
 \begin{center}
\begin{tabular}{lc}
\hline\hline
 Decay chain                                             & $S(p)$ \\ 
\hline
 $D_s^+[S] \rightarrow V_1V_2$                           & $\tilde{t}^{(1)\mu}(V_1) \; \tilde{t}^{(1)}_\mu(V_2)$ \\ 
 $D_s^+[P] \rightarrow V_1V_2$                           & $\epsilon_{\mu\nu\lambda\sigma}p^\mu(D_s^+) \; \tilde{T}^{(1)\nu}(D_s^+) \; \tilde{t}^{(1)\lambda}(V_1) \; \tilde{t}^{(1)\sigma}(V_2)$\\ 
 $D_s^+[D] \rightarrow V_1V_2$                           & $\tilde{T}^{(2)\mu\nu}(D_s^+) \; \tilde{t}^{(1)}_\mu(V_1) \; \tilde{t}^{(1)}_\nu(V_2)$\\ 
 $D_s^+ \rightarrow AP_1$,~$A[S] \rightarrow VP_2$       & $\tilde{T}^{(1)\mu}(D_s^+) \; P^{(1)}_{\mu\nu}(A) \; \tilde{t}^{(1)\nu}{(V)}$ \\
 $D_s^+ \rightarrow AP_1$,~$A[D] \rightarrow VP_2$       & $\tilde{T}^{(1)\mu}(D_s^+) \; \tilde{t}^{(2)}_{\mu\nu}(A)  \; \tilde{t}^{(1)\nu}(V)$ \\
 $D_s^+ \rightarrow AP_1$,~$A \rightarrow SP_2$          & $\tilde{T}^{(1)\mu}(D_s^+) \; \tilde{t}^{(1)}_\mu(A)$ \\
 $D_s^+ \rightarrow VS$                                  & $\tilde{T}^{(1)\mu}(D_s^+) \; \tilde{t}^{(1)}_\mu(V)$ \\
 $D_s^+ \rightarrow V_1P_1$,~$V_1\rightarrow V_2P_2$     & $\epsilon_{\mu\nu\lambda\sigma} \; p^\mu_{V_1} r^\nu_{V_1} \; p^\lambda_{P_1} \; r^\sigma_{V_2}$\\
 $D_s^+ \rightarrow PP_1$,~$P\rightarrow VP_2$           & $p^\mu(P_2) \; \tilde{t}^{(1)}_\mu(V)$\\
 $D_s^+ \rightarrow PP_1$,~$P\rightarrow SP_2$           & 1\\
 $D_s^+ \rightarrow SS$                                  & 1\\
\hline\hline
\end{tabular}
\end{center}
\end{table*}

\subsubsection{Blatt-Weisskopf Barrier Factors}
The Blatt-Weisskopf barrier $X(p_j)$~\cite{Zou2003,shilei1,shilei2,shijiu} is a barrier function for a two-body decay 
process, $a \rightarrow bc$. The Blatt-Weisskopf barrier depends on the angular momenta 
and the momenta of the final state particles in the rest frame of the 
parent particle. The definition is given by
\begin{align}  
\begin{aligned}
  X_{L=0}(q)&=1,\\   
  X_{L=1}(q)&=\sqrt{\frac{2}{q^2+(1/R)^2}},\\
  X_{L=2}(q)&=\sqrt{\frac{13}{q^4+3q^2(1/R)^2+9(1/R)^4}}\,.
\end{aligned} 
\end{align}
where $L$ denotes the orbital angular momentum; 
$R$ is the effective radius of the barrier; the values of $R$ used in this analysis are taken to be $3.0$~GeV$^{-1}$ and $5.0$~GeV$^{-1}$ 
for intermediate resonances and the $D_s^+$ meson, respectively~\cite{jiushily}; and $q$ is the magnitude of the momenta of the final-state particles in the rest system of the parent particle.

For a process $a\rightarrow bc$, $s_i=E^2_i-p_i^2$ is defined, where $i$ denotes $a,b,c$ and $E_i$, $p_i$ are the particle's energy and momentum, such that
\begin{eqnarray}
\begin{aligned}
q^2 = \frac{(s_a+s_b-s_c)^2}{4s_a}-s_b\,.
\end{aligned}  
\end{eqnarray}
\subsubsection{Propagators}
The relativistic Breit-Wigner (RBW) function is used as the propagator
for the resonances $\phi$, $\bar{K}^{*0}$, $K^{*\pm}$, $\bar{K}_1^0(1270)$, 
$\bar{K}_1^0(1400)$, $f_1(1510)$, $f_1(1420)$, and $\eta(1475)$, and 
their masses and widths are fixed to their PDG values~\cite{PDG}, as
listed in Table~\ref{pdgwhywhat}.
\begin{table}[htp]
  \centering
  \caption{The masses and widths of intermediate resonances used in this analysis.}
  \vspace{2mm} 
  \label{pdgwhywhat}
\renewcommand\arraystretch{1.2}
  \begin{tabular}{ccc}
\hline\hline
   Resonance            &      Mass (MeV/$c^2$)     & Width (MeV) \\ \hline
   $\phi$               &    $1019.461\pm0.016$     &  $4.249\pm0.013$  \\ 
   $\rho^+$             &    $775.11\pm0.34$        &  $149.1\pm0.8$    \\ 
   $\bar{K}^{*0}$       &    $895.55\pm0.20$        &  $47.3\pm0.5$     \\ 
   $K^{*\pm}$           &    $891.66\pm0.26$        &  $50.8\pm0.9$     \\ 
   $\bar{K}_1^0(1270)$  &    $1272\pm7$             &  $87\pm7$         \\ 
   $\bar{K}_1^0(1400)$  &    $1403\pm7$             &  $174\pm13$       \\ 
   $f_1(1420)$          &    $1426.3\pm0.9$         &  $54.5\pm2.6$     \\ 
   $\eta(1475)$         &    $1475\pm4$             &  $90\pm9$         \\ 
   $a_0^0(980)$         &    $990\pm1$         &  $g_{\eta\pi(K\bar{K})}$~(see text)  \\ 
\hline\hline
  \end{tabular}
\end{table}

The RBW function is given by 
\begin{eqnarray} \begin{aligned}
P(m)=\frac{1}{(m^2_0-m^2)-im_0\Gamma(m)}\,, 
\end{aligned}
\end{eqnarray} where $m=\sqrt{E^2-p^2}$ and $m_0$ is the nominal mass
of the resonance, and $\Gamma(m)$ is given by
\begin{eqnarray}
\Gamma(m)=\Gamma_0\left(\frac{q}{q_0}\right)^{2L+1}\left(\frac{m_0}{m}\right)\left(\frac{X_L(q)}{X_L(q_0)}\right)^2\,,
\end{eqnarray}
where $q_0$ indicates the value of $q$ when $s_a=m^2_0$.

Considering the obvious mass deviation,
the mass and width of $\bar{K}_1^0(1270)$ are set to the average values
(shown in Table~\ref{pdgwhywhat}) without including the results from Belle~\cite{belleaaa}.

The $\rho^+$ meson is parameterized with the Gounaris-Sakurai lineshape
(GS)~\cite{PhysRevLett.21.244}, which is given by
\begin{eqnarray}  
\begin{aligned}
P_{\text{GS}}(m)=\frac{1+d\frac{\Gamma_0}{m_0}}{(m_0^2-m^2)+f(m)-im_0\Gamma(m)}\,. 
\end{aligned}
\end{eqnarray}

The function $f(m)$ is given by
\begin{align}  
\begin{aligned}
f(m)=&\Gamma_0\frac{m_0^2}{q_0^3}\times\Bigg[q^2\Big (h(m)-h(m_0)\Big )\\
&+(m_0^2-m^2)q_0^2\left.\frac{dh}{d(m^2)}\right|_{m_0^2}\Bigg]\,,     
\end{aligned}
\end{align}
where
\begin{align}
h(m)&=\frac{2q}{\pi m}\ln\left(\frac{m+2q}{2m_{\pi}}\right)\,, 
\\
\left.\frac{dh}{d(m^2)}\right|_{m_0^2}&=h(m_0)\left[(8q_0^2)^{-1}-(2m_0^2)^{-1}\right]+(2\pi m_0^2)^{-1}\,,
\end{align}
and $m_{\pi}$ is the charged pion mass.

The normalization condition at $P_{\text{GS}}(0)$ fixes the parameter $d=f(0)/(\Gamma_0 m_0)$. It is found to be
\begin{eqnarray}  
\begin{aligned}
d=\frac{3m^2_\pi}{\pi q_0^2}\ln\left(\frac{m_0+2q_0}{2m_\pi}\right)+\frac{m_0}{2\pi q_0}-\frac{m^2_\pi m_0}{\pi q^3_0}\,.  
\end{aligned}    
\end{eqnarray}

The $a_0(980)$ meson lineshape is parameterized by the Flatt\'e formula~\cite{flatte},
\begin{eqnarray}  
\begin{aligned}
P_{a_0(980)}=\frac{1}{(m_0^2-s_a)-i(g_{\eta\pi}^2\rho_{\eta\pi}+g^2_{K\bar{K}}\rho_{K\bar{K}})}\,,  
\end{aligned}    
\end{eqnarray}
where $m_0$ is the mass of $a_0(980)$ and
$g_{\eta\pi(K\bar{K})}^2$ is the coupling constant. These parameters
are fixed at the values given in Ref.~\cite{a0980}, in which
$m_0=(0.990\pm0.001)$GeV/$c^2$,
$g_{\eta\pi}^2=(0.341\pm0.004)$GeV$^2$/$c^4$ and
$g_{K\bar{K}}^2=(0.892\pm0.022)g_{\eta\pi}^2$.  The
$\rho_{\eta\pi(K\bar{K})}$ is the PHSP factor and is given by
$\rho_{\eta\pi(K\bar{K})}=2q/\sqrt{s_a}$.

The $K\pi$ ${\rm S{\text -}wave}$ is
modeled by a parameterization from scattering
data~\cite{PhysRevD.98.112012}, which is built from a Breit-Wigner
shape for the $K^{*}_0(1430)$ resonance combined with an effective
range parameterization for the non-resonant component with a phase
shift given by
\begin{eqnarray}\label{eq:kpi_swave_formfactor}                                                                                                                                    
\begin{aligned}
  A(m)=F\sin\delta_F e^{i\delta_F}+R\sin\delta_R e^{i\delta_R}e^{i2\delta_F}\,,                                                                         
\end{aligned}
\end{eqnarray}

with
\begin{eqnarray}                                                                                                                                                                              
\begin{aligned}
  \delta_F&=\phi_F+\cot^{-1}\left[\frac{1}{aq}+\frac{rq}{2}\right]\nonumber\,,\\                                                                                                                                     
  \delta_R&=\phi_R+\tan^{-1}\left[\frac{M\Gamma(m_{K\pi})}{M^2-m^2_{K\pi}}\right]\,,\nonumber                                                           
\end{aligned}
\end{eqnarray}
where $a$ and $r$ are the scattering length and effective interaction
length, respectively.  The parameters $F(\phi_F)$ and $R(\phi_R)$ are
the magnitude (phase) for the non-resonant state and resonance terms,
respectively.  The parameters $M$, $F$, $\phi_F$, $R$, $\phi_R$, $a$
and $r$ are fixed to the results of the $D^0\rightarrow
K^0_S\pi^+\pi^-$ analysis by the BABAR and Belle
experiments~\cite{PhysRevD.98.112012} and are given in
Table~\ref{tab:Kpi_swave_par}.
\begin{table}[ht]
 \caption{The $K\pi$ ${\rm S{\text -}wave}$ parameters, obtained from
   a fit to the $D^0\rightarrow K^0_S\pi^+\pi^-$ Dalitz plot from the
   BABAR and Belle experiments~\cite{PhysRevD.98.112012}. The first
   uncertainties are statistical and the second
   systematic.}\label{tab:Kpi_swave_par}
 \begin{center}
 \centering
\renewcommand\arraystretch{1.2}
 \begin{tabular}{cc}
\hline\hline
$M$(GeV/$c^2$) &$1.441\pm 0.002$\\
$\Gamma$(GeV) &$0.193\pm 0.004$\\
$F$ &$0.96\pm 0.07$\\
$\phi_F(\rm deg)$&$0.1\pm0.3$\\
$R$&1 (fixed)\\
$\phi_R(\rm deg)$&$-109.7\pm 2.6$\\
$a$&$0.113\pm 0.006$\\
$r$&$-33.8\pm 1.8$\\
\hline\hline
 \end{tabular}
 \end{center}
\end{table}
\subsection{\boldmath Fit Fractions and Statistical Uncertainty}
The fit fractions of the individual components (amplitudes) are
calculated according to the fit results. In the calculation, a large
PHSP MC sample (twelve million events) with neither detector
acceptance nor resolution included is used. The fit fraction (FF) for
a component or an amplitude is defined as \begin{align} \text{FF}_n =
\frac{\int\vert c_nA_n(p_j)\vert^2R_4(p_j)dp_j}{\int\vert
M(p_j)\vert^2R_4(p_j)dp_j}\approx \frac{\sum\limits_{k=1}^{N_{\rm
g,\text{ph}}}\vert
\tilde{A}_n(p_j^k)\vert^2}{\sum\limits_{k=1}^{N_{\rm
g,\text{ph}}}\vert M(p_j^k)\vert^2}\,, \end{align} where the
integration is approximated by the PHSP MC summation at the generator
level, $\tilde{A}_n(p_j^k)$ is either the $n^{\rm th}$ amplitude
($\tilde{A}_n(p_j^k)=c_nA_n(p_j^k)$) or the $n^{\rm th}$ component of
a coherent sum of amplitudes ($\tilde{A}_n(p_j^k)=\sum
c_{n_i}A_{n_i}(p_j^k)$), and $N_{\rm g,\text{ph}}$ is the number of
PHSP MC events.

For the statistical uncertainty of FF, it is impractical to
analytically propagate the uncertainties of the FFs from those of the
magnitudes and phases. Instead, the variables are randomly perturbed within their uncertainties 
obtained from the fit, and the FFs are calculated to determine the statistical uncertainties. 
The distribution of each FF is fitted with a Gaussian function, and the 
width is the statistical uncertainty of the FF.

\subsection{\boldmath Results of the Amplitude Analysis}\label{kakuka}
The amplitude of the
$D_s^+[S]\to \phi\rho^+$ decay is expected to have the largest
FF. Thus, this amplitude is chosen as the reference (its phase is
fixed to 0, and the magnitude is fixed to 1). The notation $[S]$
denotes a relative orbital angular momentum $L=0$ between daughters in
a decay, and similarly for $[P]$ ($L=1$), $[D]$ ($L=2$). In addition,
 some necessary physical relations are fixed, as shown in
Appendix~\ref{relations}.

The fit to the data is initially performed with a baseline model including the
amplitudes of $D_s^+\to \phi\rho^+$, $D_s^+\to \bar{K}^{*0}K^{*+}$,
$D_s^+\to \bar{K}_1^0(1270)K^+$ ($\bar{K}_1^0(1270)\to K^-\rho^+$ and
$K^*\pi$) and $D_s^+\to \bar{K}_1^0(1400)K^+$ ($\bar{K}_1^0(1400)\to
K^{*}\pi$) decays, as the $\phi$, $\rho^+$, $\bar{K}^{*0}$, $K^{*+}$,
$K^{*-}$, $\bar{K}_1^0(1270)$, and $\bar{K}_1^0(1400)$ resonances are
clearly observed in the corresponding invariant mass spectra. The
statistical significances (SSs) of the above amplitudes, which are
determined from the changes in log-likelihood and the numbers of degrees
of freedom (NDOF) when the fits are performed with and without the
amplitude included, are all much larger than $4\sigma$.

Starting from the baseline model above, the amplitudes involving
$f_1(1420)$, $f_1(1510)$, $\eta(1405)$, and $\eta(1475)$ resonances are added 
to improve the fit quality of the $K^-K^+\pi^0$ invariant mass
spectrum. The amplitudes with significances larger than $4\sigma$ are retained 
for the next iteration. The amplitudes of $D_s^+\to f_1(1420)\pi^+$
($f_1(1420)\to K^*K$) and $D_s^+\to \eta(1475)\pi^+$ ($\eta(1475)\to
a_0^0(980)\pi^0$) decays have significances larger than $5\sigma$, and
the amplitude of $D_s^+\to f_1(1420)\pi^+$ ($f_1(1420)\to
a_0^0(980)\pi^0$) decay improves the fit of the $K^-K^+\pi^0$ mass
spectrum. Then, other amplitudes are tested, but only the $D_s^+\to
a_0^0(980) \rho^+$ decay is significant ($9\sigma$). Finally, eighteen 
amplitudes are retained in the nominal fit, which are categorized into 
nine processes, as shown in Table~\ref{ninedecay}.
The amplitudes of the nominal fit are listed in Table~\ref{fitresult-table}.
Other possible processes with significance less than $3\sigma$ are listed in
Appendix~\ref{othertest}.
\begin{table}[htp]
\caption{The nine components in the nominal amplitude model.}
  \label{ninedecay}
  \centering
\renewcommand\arraystretch{1.2}
  \begin{tabular}{l}
\hline\hline
$D_s^+\to \phi\rho^+$ \\
$D_s^+\to \bar{K}^{*0}K^{*+}$\\
$D_s^+\to a_0^0(980) \rho^+$\\
$D_s^+\to \bar{K}_1^0(1270)K^+(\bar{K}_1^0(1270)\to K^-\rho^+)$\\
$D_s^+\to \bar{K}_1^0(1270)K^+(\bar{K}_1^0(1270)\to K^{*}\pi)$ \\
$D_s^+\to \bar{K}_1^0(1400) K^+(\bar{K}_1^0(1400)\to K^{*}\pi)$\\
$D_s^+\to f_1(1420)\pi^+(f_1(1420)\to K^{*\mp}K^{\pm})$\\
$D_s^+\to f_1(1420)\pi^+(f_1(1420)\to a_0^0(980)\pi^0)$\\
$D_s^+\to \eta(1475)\pi^+(\eta(1475)\to a_0^0(980)\pi^0)$\\ 
\hline\hline
  \end{tabular}
\end{table}

\begin{figure*}[htbp]
\centering
    \mbox{
    \begin{overpic}[width=5.7cm,height=4.6cm,angle=0]{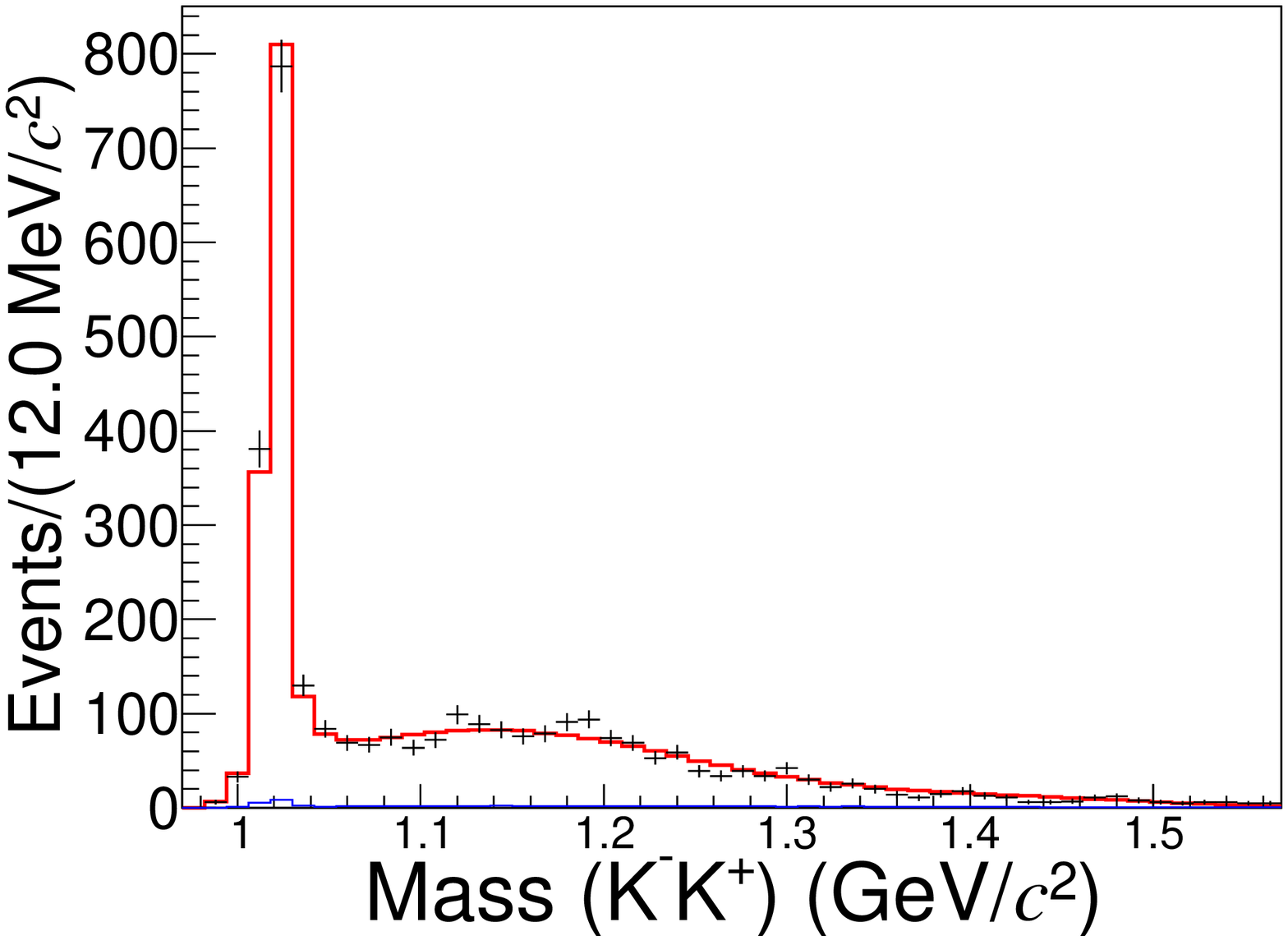}
    \put(80,50){{ (a) }}
    \end{overpic}
    \begin{overpic}[width=5.7cm,height=4.6cm,angle=0]{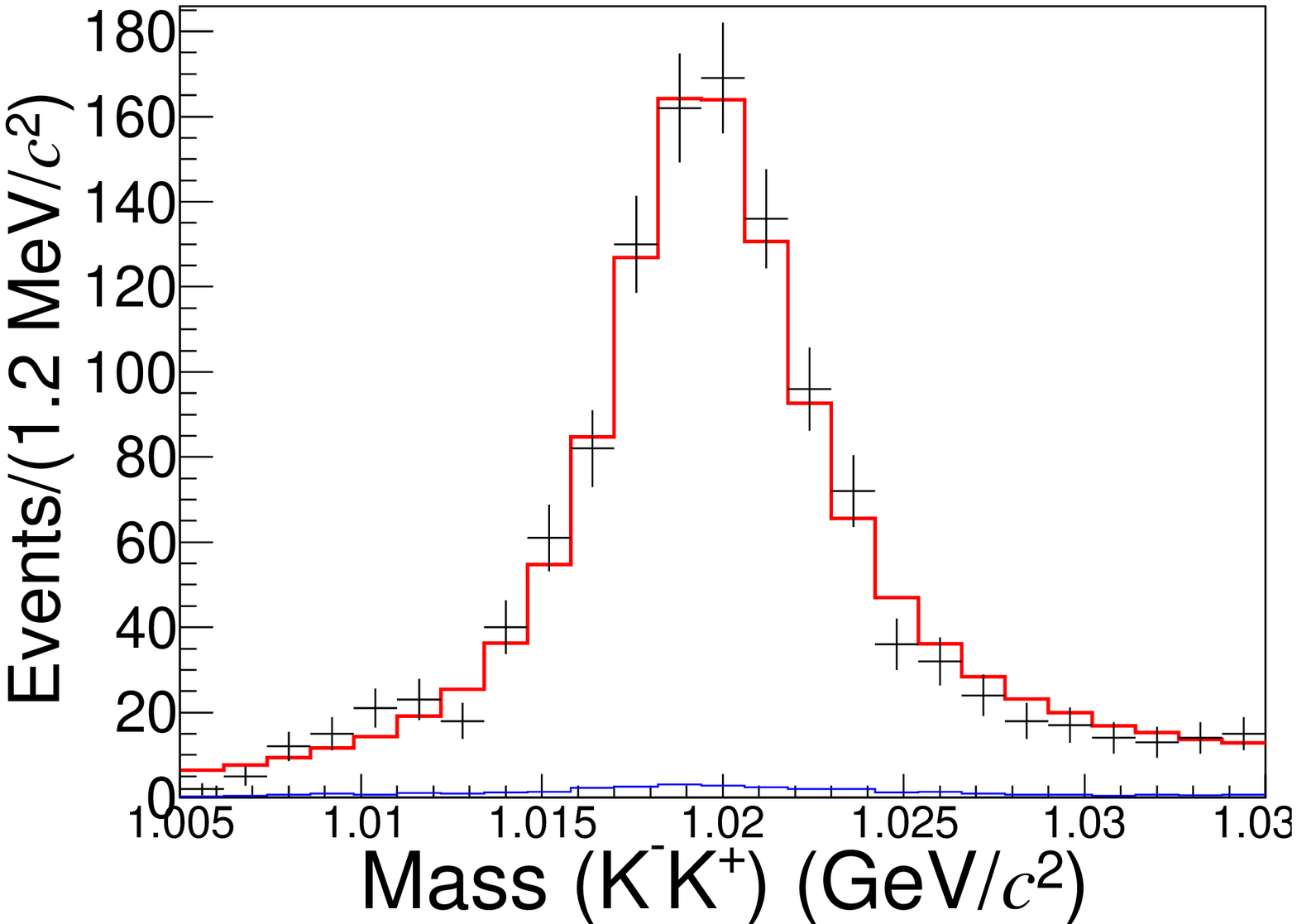}
    \put(80,50){{ (b) }}
    \end{overpic}
    \begin{overpic}[width=5.7cm,height=4.6cm,angle=0]{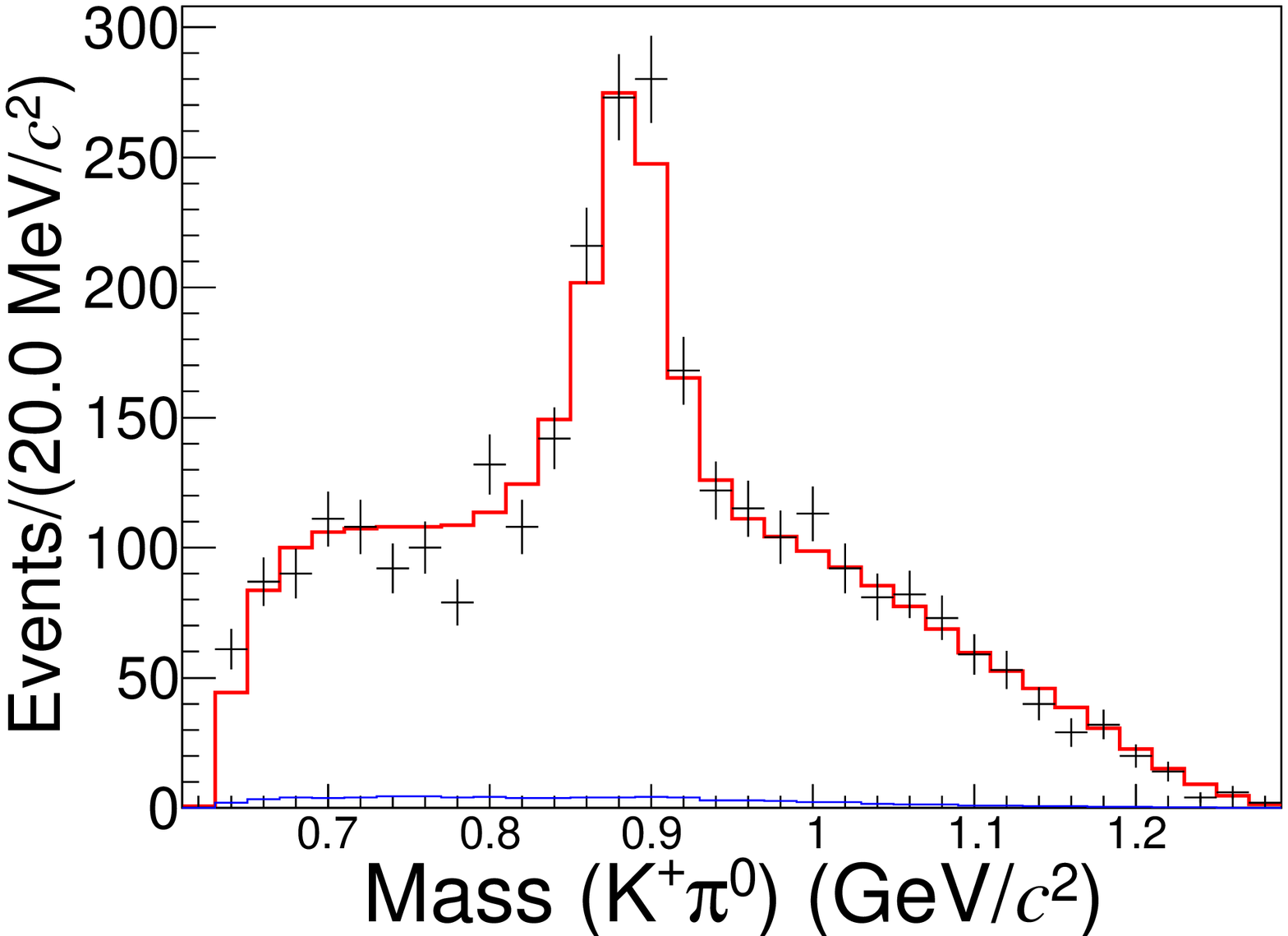}
    \put(80,50){{ (c) }}
    \end{overpic}    }
    \mbox{
    \begin{overpic}[width=5.7cm,height=4.6cm,angle=0]{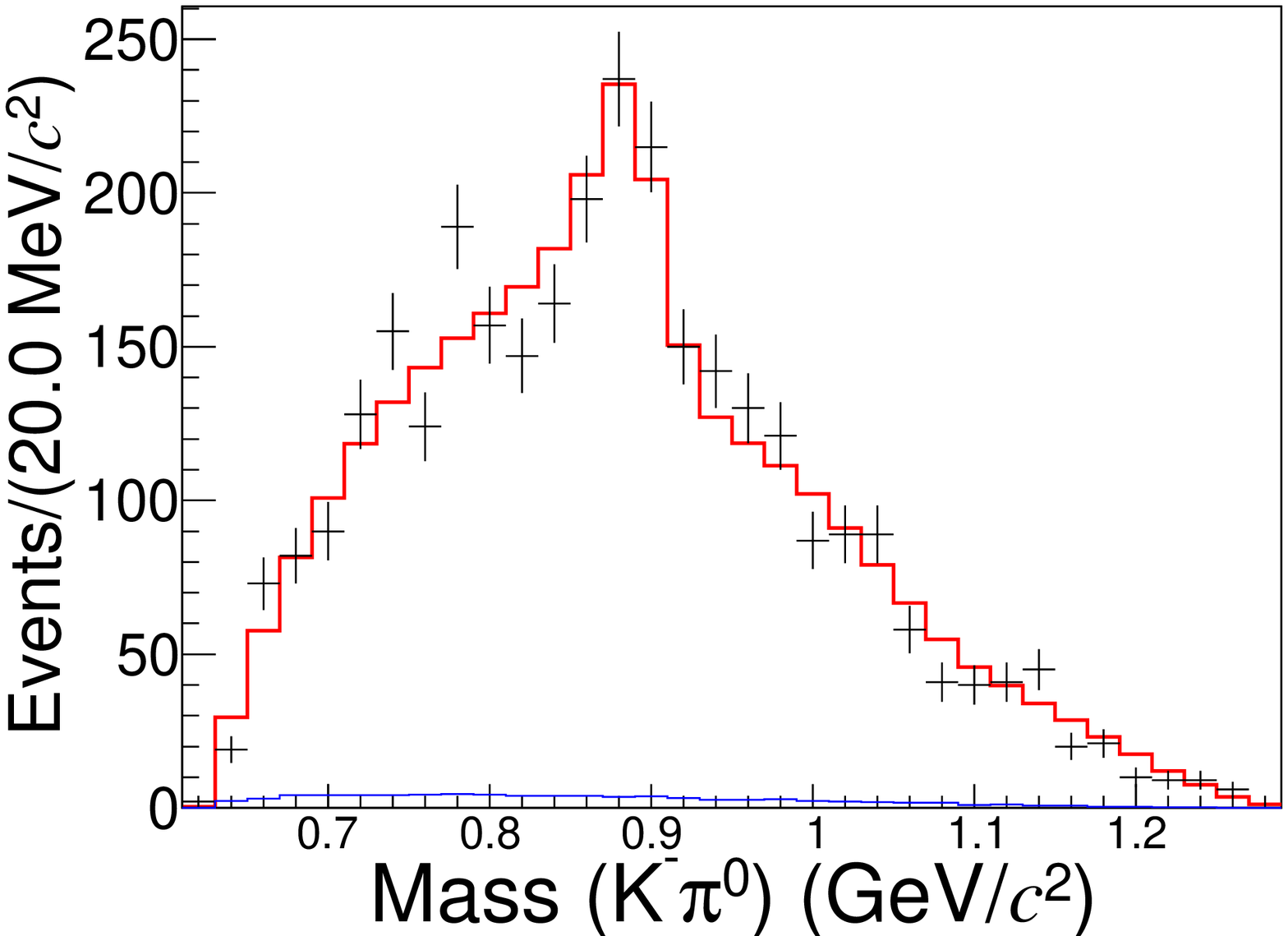}
    \put(80,50){{ (d) }}
    \end{overpic}
    \begin{overpic}[width=5.7cm,height=4.6cm,angle=0]{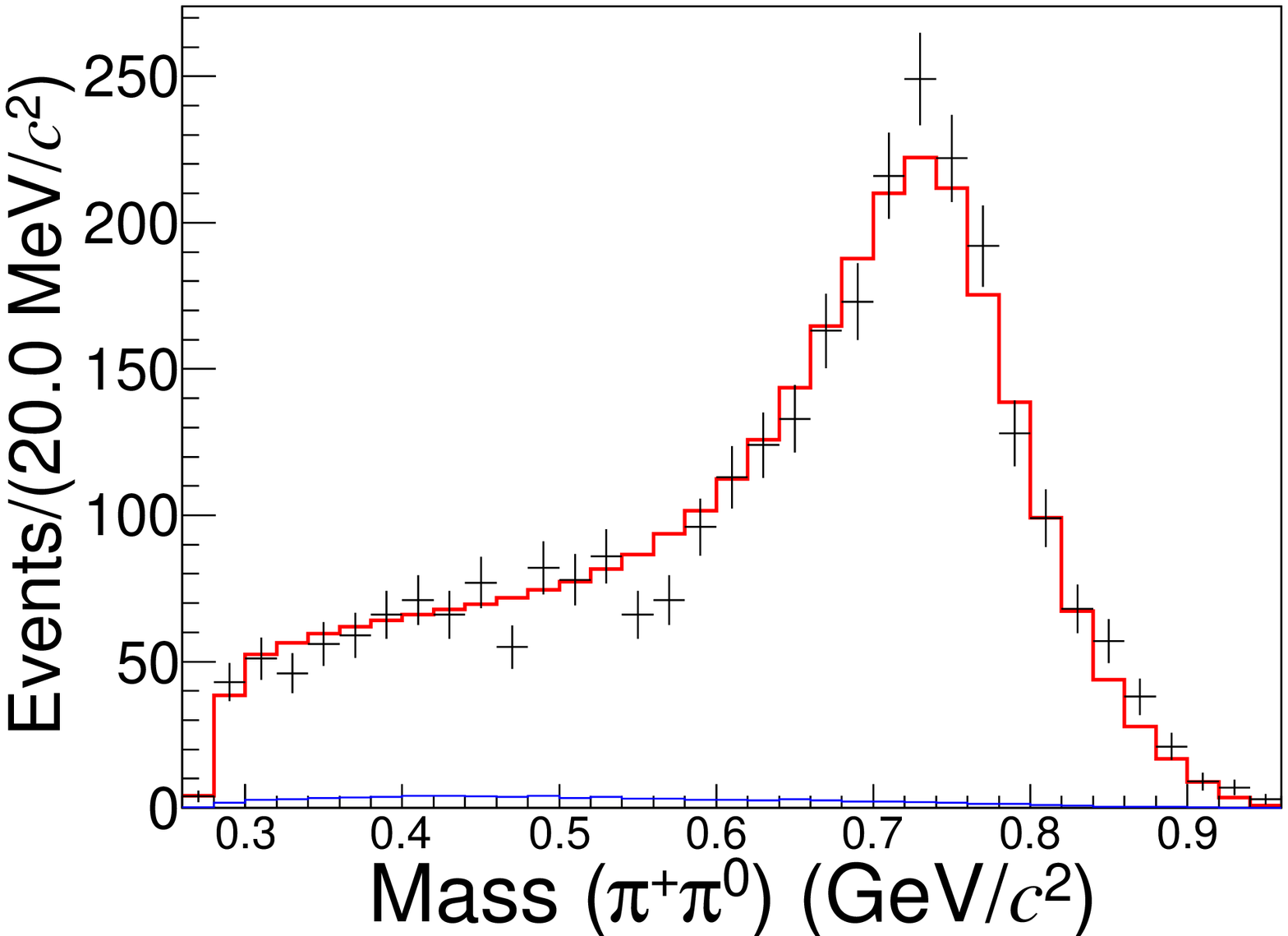}
    \put(80,50){{ (e) }}
    \end{overpic}
    \begin{overpic}[width=5.7cm,height=4.6cm,angle=0]{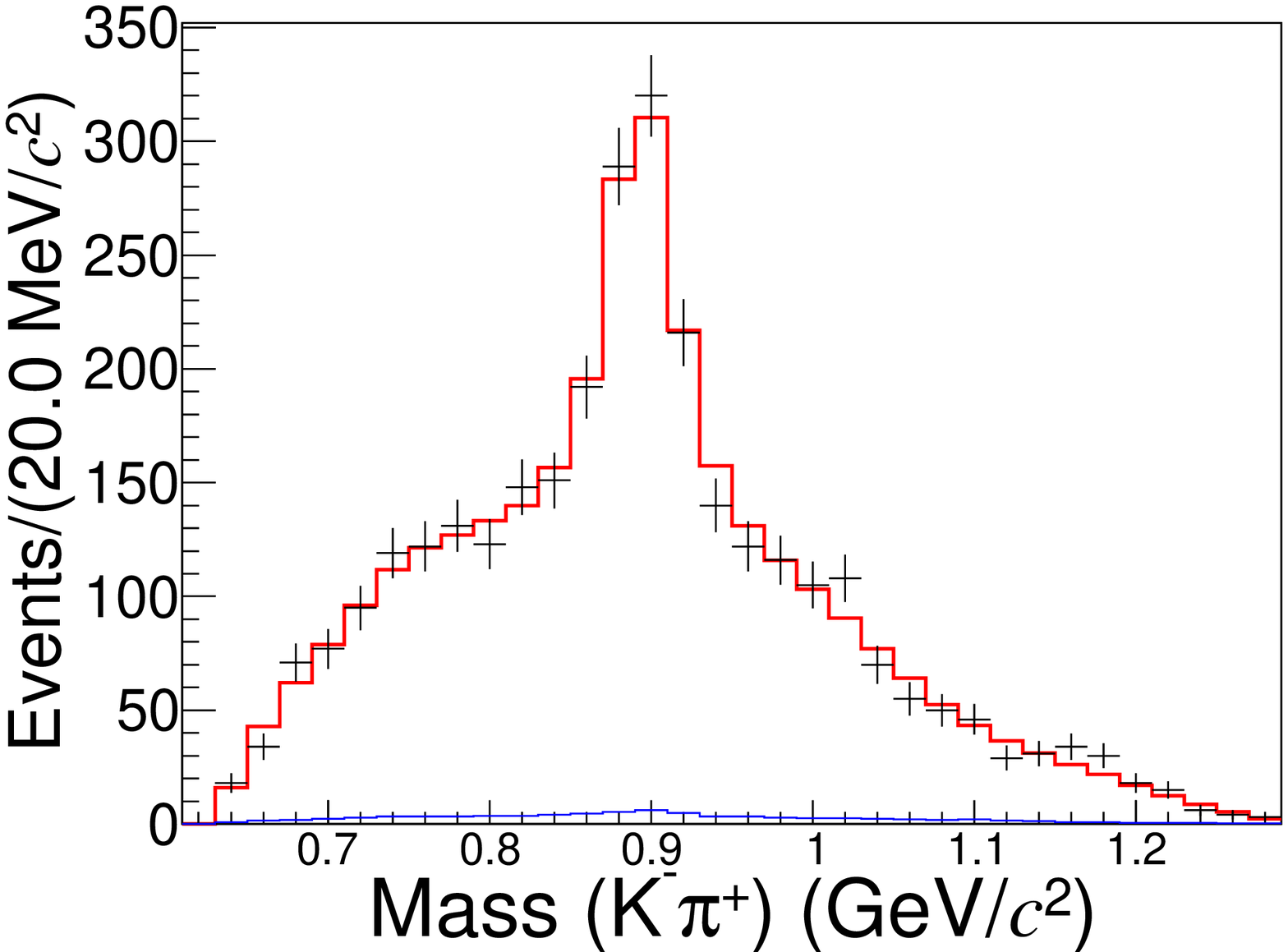}
    \put(80,50){{ (f) }}
    \end{overpic}    }
    \mbox{
    \begin{overpic}[width=5.7cm,height=4.6cm,angle=0]{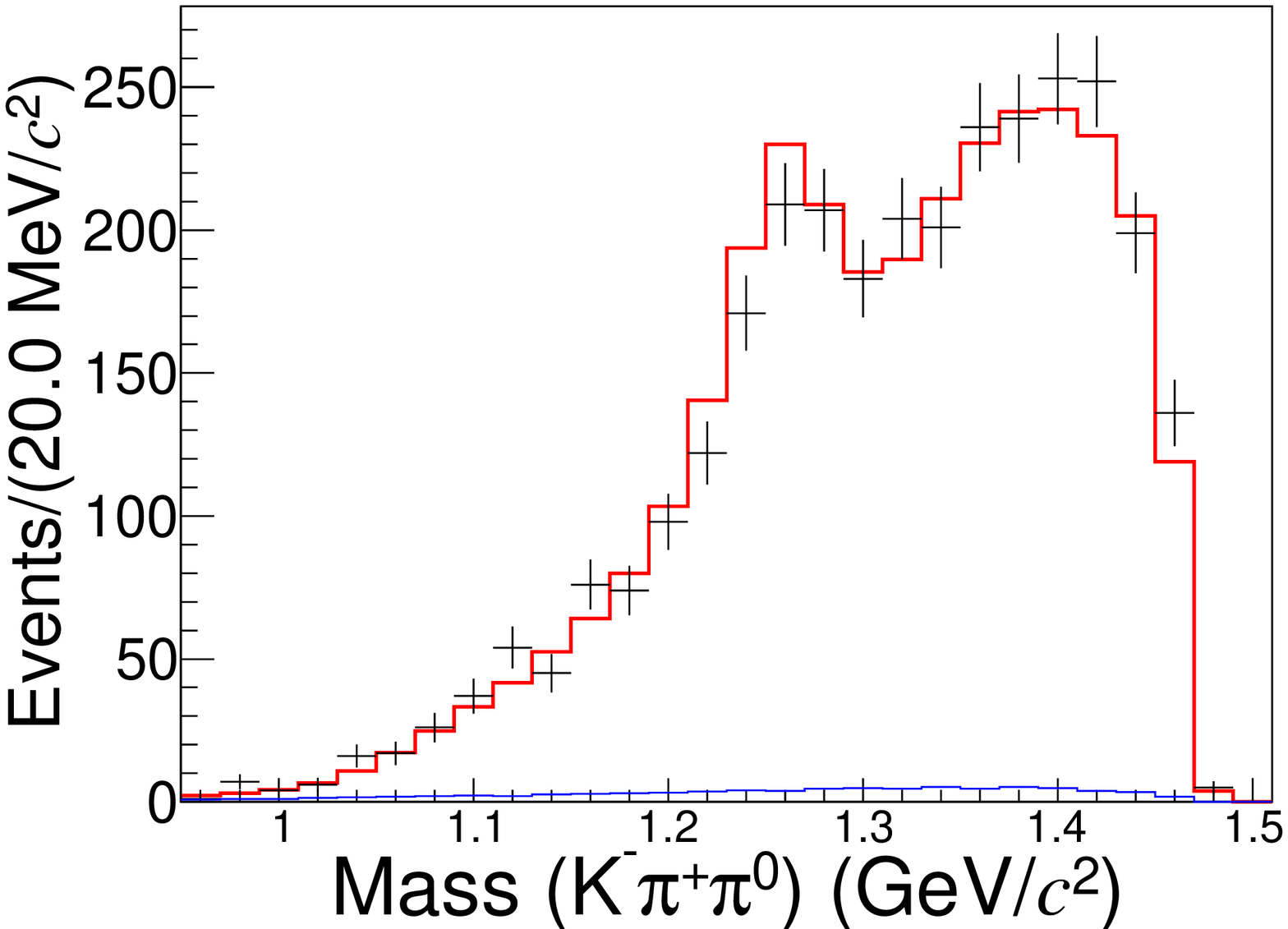}
    \put(30,50){{ (g) }}
    \end{overpic}
    \begin{overpic}[width=5.7cm,height=4.6cm,angle=0]{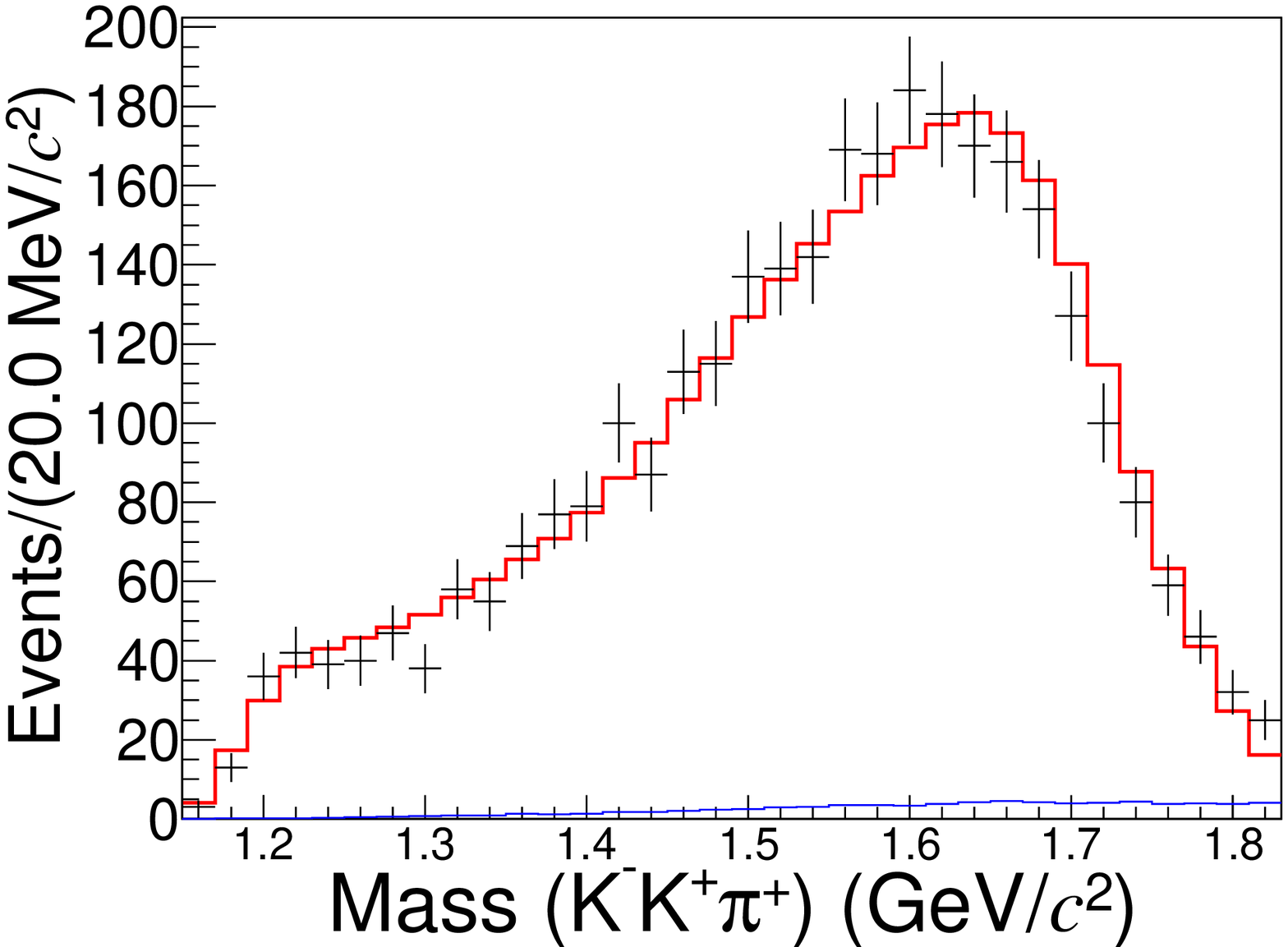}
    \put(30,50){{ (h) }}
    \end{overpic}
    \begin{overpic}[width=5.7cm,height=4.6cm,angle=0]{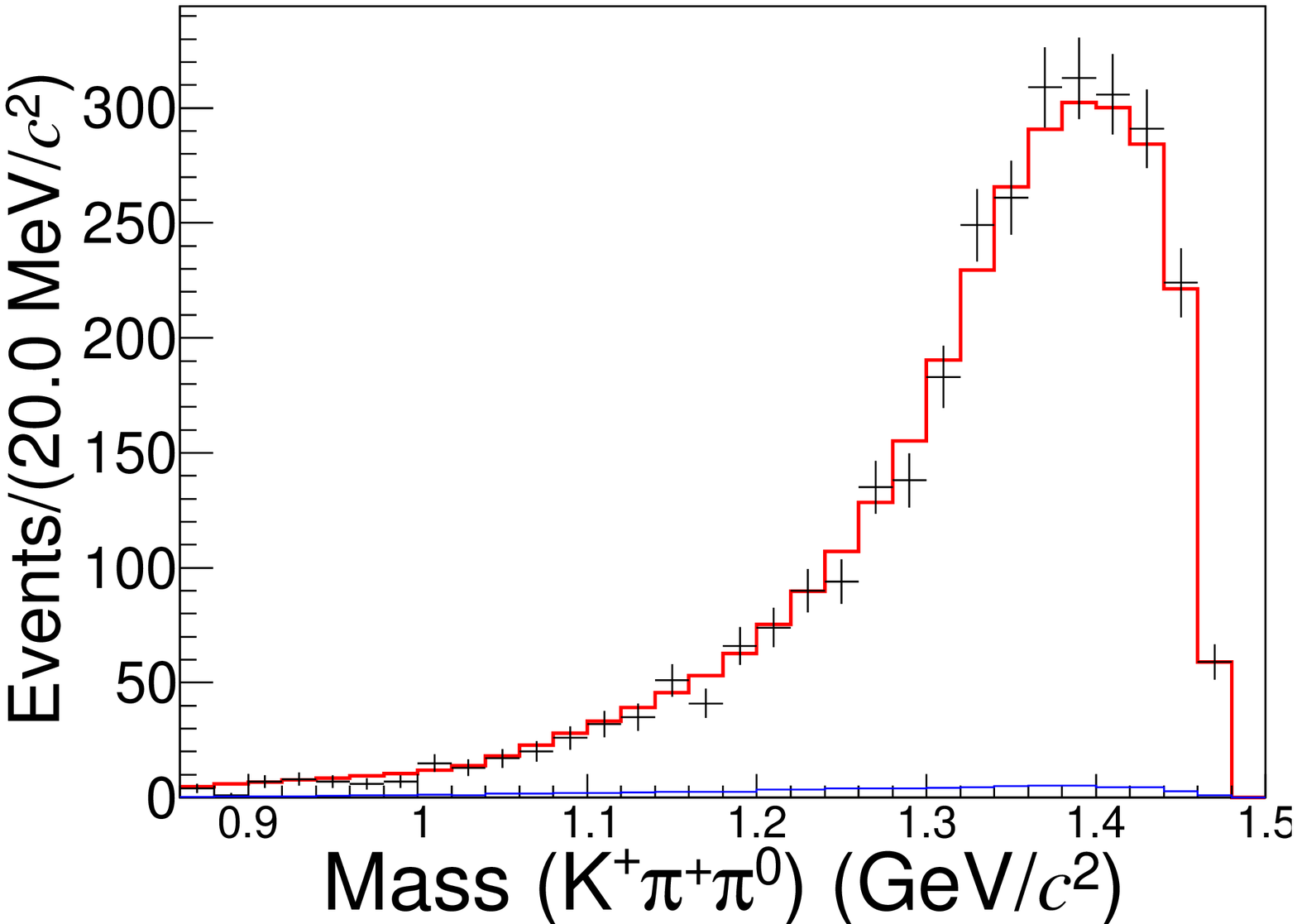}
    \put(30,50){{ (i) }}
    \end{overpic}    }
    \mbox{
    \begin{overpic}[width=5.7cm,height=4.6cm,angle=0]{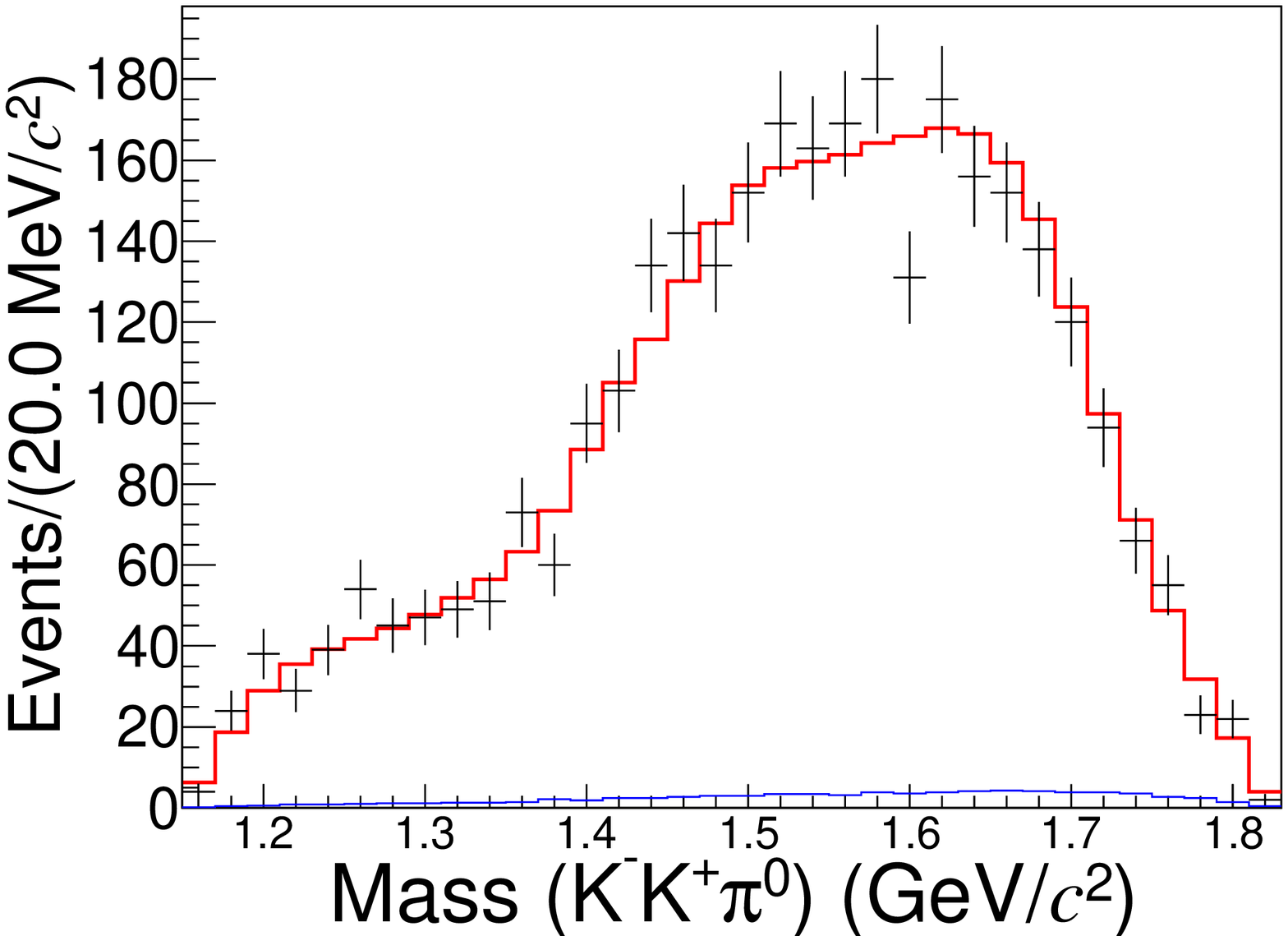}
    \put(30,50){{ (j) }}
    \end{overpic}    }
\caption{Invariant mass distributions of (a) $K^-K^+$, (c) $K^+\pi^0$, (d) $K^-\pi^0$, (e) $\pi^+\pi^0$, (f) $K^-\pi^+$, (g) $K^-\pi^+\pi^0$, (h) $K^-K^+\pi^+$, (i) $K^+\pi^+\pi^0$, and (j) $K^-K^+\pi^0$, where the black points with error bars are data, and the red histograms are for the fit projections. The blue histograms are the backgrounds. Plot (b) shows the $\phi$ mass region of plot (a) with an expanded scale.}\label{pwafitplot333} \end{figure*}

The fit results with phases, FFs and SSs for each amplitude are shown
in Table~\ref{fitresult-table}. The ratio $\frac{{\cal B}(D_s^+\to
  \bar{K}_1^0(1270)K^+,\bar{K}_1^0(1270)\to K^{*-}\pi^+)}{{\cal
    B}(D_s^+\to \bar{K}_1^0(1270)K^+,\bar{K}_1^0(1270)\to K^-\rho^+)}$
is determined to be $0.33 \pm 0.05_{\rm stat.} \pm 0.06_{\rm syst.}$ in this analysis, accounting
for correlations.
The fit projections of three data samples on the invariant masses are shown in
Fig.~\ref{pwafitplot333}.

\begin{table*}[htp]
 \centering
  \caption{Phase, FF, and SS for the different amplitudes, labeled as
    I, II..., XIV. Groups of related amplitudes are separated by horizontal lines. The last row of each group
    gives the total fit fraction of the above components with interferences considered. The amplitudes VIII, IX,
    X, and XII are constructed by two sub-amplitudes with fixed
    relations (see Appendix~\ref{relations}). The $\rho^+$ resonance
    decays to $\pi^+\pi^0$. The $\phi$ and $a_0^0(980)$ resonances
    decay to $K^-K^+$. The $\bar{K}^{*0}$ resonance decays to
    $K^-\pi^+$, and the $K^{*\pm}$ resonance decays to $K^{\pm}\pi^0$.
    $K^{*}\pi$ indicates $\bar{K}^{*0}\pi^0$ and $K^{*-}\pi^+$. The first and second uncertainties are statistical and systematic, respectively.}
  \vspace{2mm} 
  \label{fitresult-table}
\renewcommand\arraystretch{1.2}
  \begin{tabular}{ccccc}
\hline\hline
Label&                                         Amplitude                                            &  Phase ($\phi_{n}$)        &            FF (\%)          &  SS $(\sigma)$\\\hline
  I &   $D_s^+[S]\to \phi\rho^+$                                                                    & 0.0 (fixed)                &$    38.68\pm1.42\pm2.17$    &  $\textgreater20$ \\   
  II&   $D_s^+[P]\to \phi\rho^+$                                                                    & $ -1.46\pm0.05\pm0.02$     &$ \, ~9.64\pm0.84\pm0.30$    &  17.1  \\ 
 III&   $D_s^+[D]\to \phi\rho^+$                                                                    & $  1.46\pm0.07\pm0.04$     &$ \, ~3.36\pm0.75\pm0.27$    &  4.5  \\  
    &   $D_s^+\to \phi\rho^+$                                                                       &     $\cdots$               &$    50.81\pm1.01\pm2.20$    &   $\cdots$  \\\hline  
  IV&   $D_s^+[S]\to \bar{K}^{*0}K^{*+}$                                                            & $ -2.15\pm0.06\pm0.05$     &$    16.32\pm0.95\pm0.33$    &  $\textgreater20$  \\ 
   V&   $D_s^+[P]\to \bar{K}^{*0}K^{*+}$                                                            & $ -0.52\pm0.07\pm0.04$     &$ \, ~6.87\pm0.55\pm0.26$    &  16.1  \\ 
  VI&   $D_s^+[D]\to \bar{K}^{*0}K^{*+}$                                                            & $ -1.57\pm0.08\pm0.03$     &$ \, ~3.34\pm0.55\pm0.18$    &  12.1  \\ 
    &   $D_s^+\to \bar{K}^{*0}K^{*+}$                                                               &     $\cdots$               &$    23.15\pm0.89\pm0.74$    &   $\cdots$  \\\hline 
 VII&   $D_s^+\to \bar{K}_1^0(1270)K^+$,~$\bar{K}_1^0(1270)\to K^-\rho^+$                           & $  1.87\pm0.08\pm0.17$     &$    10.44\pm0.81\pm0.73$    &  $\textgreater20$  \\\hline 
    &   \footnotesize{$D_s^+\to \bar{K}_1^0(1270)K^+$,~$\bar{K}_1^0(1270)[S]\to \bar{K}^{*0}\pi^0$} &     $\cdots$               &$ \, ~1.40\pm0.26\pm0.17$    &   $\cdots$ \\  
    &   \footnotesize{$D_s^+\to \bar{K}_1^0(1270)K^+$,~$\bar{K}_1^0(1270)[S]\to K^{*-}\pi^+$}       &     $\cdots$               &$ \, ~2.60\pm0.48\pm0.31$    &   $\cdots$ \\  
VIII&   $D_s^+\to \bar{K}_1^0(1270)K^+$,~$\bar{K}_1^0(1270)[S]\to K^{*}\pi$                         & $ -0.25\pm0.11\pm0.12$     &$ \, ~3.88\pm0.71\pm0.45$    &  10.8  \\
    &   \footnotesize{$D_s^+\to \bar{K}_1^0(1270)K^+$,~$\bar{K}_1^0(1270)[D]\to \bar{K}^{*0}\pi^0$} &     $\cdots$               &$ \, ~0.45\pm0.11\pm0.10$    &   $\cdots$  \\  
    &   \footnotesize{$D_s^+\to \bar{K}_1^0(1270)K^+$,~$\bar{K}_1^0(1270)[D]\to K^{*-}\pi^+$}       &     $\cdots$               &$ \, ~0.86\pm0.20\pm0.17$    &   $\cdots$  \\  
  IX&   $D_s^+\to \bar{K}_1^0(1270)K^+$,~$\bar{K}_1^0(1270)[D]\to K^{*}\pi$                         & $  1.52\pm0.11\pm0.15$     &$ \, ~1.34\pm0.31\pm0.27$    &  8.3  \\
    &   $D_s^+\to \bar{K}_1^0(1270)K^+$,~$\bar{K}_1^0(1270)\to K^{*}\pi$                            &     $\cdots$               &$ \,~ 5.43\pm0.69\pm0.76$    &   $\cdots$  \\\hline  
    &   \footnotesize{$D_s^+\to \bar{K}_1^0(1400)K^+$,~$\bar{K}_1^0(1400)[S]\to \bar{K}^{*0}\pi^0$} &     $\cdots$               &$ \, ~2.90\pm0.39\pm0.44$    &   $\cdots$  \\ 
    &   \footnotesize{$D_s^+\to \bar{K}_1^0(1400)K^+$,~$\bar{K}_1^0(1400)[S]\to K^{*-}\pi^+$}       &     $\cdots$               &$ \, ~5.37\pm0.73\pm0.82$    &   $\cdots$  \\ 
   X&   $D_s^+\to \bar{K}_1^0(1400)K^+$,~$\bar{K}_1^0(1400)[S]\to K^{*}\pi$                         & $ -0.92\pm0.07\pm0.05$     &$ \, ~8.03\pm1.09\pm1.22$    &  13.1  \\\hline 
  XI&   $D_s^+\to a_0^0(980) \rho^+$                                                                & $  2.15\pm0.08\pm0.08$     &$ \, ~3.46\pm0.58\pm0.61$    &  9.9  \\\hline  
    & \footnotesize{$D_s^+\to f_1(1420)\pi^+$,~$f_1(1420)\to K^{*-}K^+$}                            &     $\cdots$               &$ \, ~1.56\pm0.28\pm0.17$    &   $\cdots$  \\  
    & \footnotesize{$D_s^+\to f_1(1420)\pi^+$,~$f_1(1420)\to K^{*+}K^-$}                            &     $\cdots$               &$ \, ~1.56\pm0.28\pm0.17$    &   $\cdots$  \\  
 XII&   $D_s^+\to f_1(1420)\pi^+$,~$f_1(1420)\to K^{*\mp}K^{\pm}$                                   & $  2.13\pm0.08\pm0.05$     &$ \, ~2.39\pm0.43\pm0.25$    &  9.5  \\\hline  
XIII&   $D_s^+\to f_1(1420)\pi^+$,~$f_1(1420)\to a_0^0(980)\pi^0$                                   & $  2.95\pm0.13\pm0.06$     &$ \, ~0.77\pm0.27\pm0.09$    &  4.5  \\\hline  
 XIV&   $D_s^+\to \eta(1475)\pi^+$,~$\eta(1475)\to a_0^0(980)\pi^0$                                 & $  0.61\pm0.10\pm0.06$     &$ \, ~1.37\pm0.32\pm0.34$    &  8.0  \\  
\hline\hline
  \end{tabular}
\end{table*}

  \subsection{\boldmath Goodness of Fit}
The goodness of the fit is checked with five invariant masses, $M_{K^-\pi^+}$,
  $M_{\pi^+\pi^0}$, $M_{K^-K^+}$, $M_{K^-\pi^+\pi^0}$ and
  $M_{K^-K^+\pi^0}$, which are
  divided into cells of equal size. When cells contain fewer
  than 10 events, adjacent cells are
  combined until the number of events in each cell is larger than
  10. For each cell, $\chi_p=\frac{N_p-N_p^{\rm exp}}{\sqrt{N_p^{\rm
  exp}}}$ is calculated, and the goodness of the fit is given by
  $\chi^2=\sum\limits_{p=1}^{n}\chi^2_p$, where $N_p$ and $N_p^{\rm
  exp}$ are the number of the observed events and the number
determined by the fit results in the $p^{\rm th}$ cell, respectively,
and $n$ is the total number of cells. NDOF is given by $(n-1)-n_{\rm
  par}$, where $n_{\rm par}$ is the number of the free parameters in
the fit. Overall, the value of $\chi^2/{\rm NDOF}$ is determined to be $316.8/273$.

\subsection{\boldmath Systematic Uncertainties}\label{PWA_systematics}
Systematic uncertainties from the amplitude model, the background
level, and the fit bias are considered. The systematic uncertainties
of phases ($\phi$) and FFs for different amplitudes are shown in
Tables~\ref{pulldistribution} and \ref{pulldistribution2},
respectively.
\begin{itemize}
\item {\it Amplitude model}\\
The systematic uncertainties related to the amplitude model involve the masses and widths of the intermediate resonances, the line shape of the $\rho^+$ meson, the parameters of $a_0^0(980)$ resonance, and the barrier effective radii of the $D_s^+$ meson and other intermediate states.
\item[-](1) The uncertainties associated with the masses and widths of
  the intermediate resonances [$\phi$, $\rho^+$, $\bar{K}^{*0}$, $K^{*\pm}$,
  $\bar{K}_1^0(1270)$, $\bar{K}_1^0(1400)$, $f_1(1420)$, $\eta(1475)$] are
  estimated by varying the corresponding masses and widths listed in
  Table~\ref{pdgwhywhat} within $1\sigma$.
\item[-](2) For the lineshape of the $\rho^+$ meson, an alternative
  lineshape parameterization with RBW replacing GS is
  used.
\item[-](3) The coupling constants and mass of $a_0^0(980)$
  resonance are varied within the uncertainties given by
  Ref.~\cite{a0980}.
\item[-](4) The barrier effective radii ($R$) of the $D_s^+$
  meson and other intermediate states are assumed to have a uniform distribution. For
  the $D_s^+$ meson, the value of $R$ is varied between 4.0 GeV$^{-1}$ and 6.0
  GeV$^{-1}$. For the intermediate states, $R$ is varied between 
  2.0 GeV$^{-1}$ and 4.0 GeV$^{-1}$.
\end{itemize}

\begin{itemize}
\item {\it Experimental effects}
\item[-](5) These effects are related to the PID, tracking, and reconstruction
  efficiency differences between data and MC, and are reflected in the factor
  $\gamma_{\epsilon}$ in Eq.~\ref{wochache}. The PID efficiencies are
  studied using clean samples of $e^+e^-\to K^+K^-K^+K^-$,
  $K^+K^-\pi^+\pi^-$, $K^+K^-\pi^+\pi^-\pi^0$, $\pi^+\pi^-\pi^+\pi^-$
  and $\pi^+\pi^-\pi^+\pi^-\pi^0$ decays, while two clean samples of
  the continuum process $K^+K^-\pi^+\pi^-$ and the $e^+e^-\to K^+K^-\pi^+\pi^-\pi^0$ decay are used for the studies of the tracking
  efficiencies and the $\pi^0$ reconstruction efficiency, respectively. These efficiencies are also used in the BF measurement
  (Section~\ref{bfzalame}). The PID, tracking, and reconstruction systematic
  uncertainties are taken as the efficiency differences between data
  and MC simulation. The uncertainties associated with
  $\gamma_{\epsilon}$ are obtained by performing alternative amplitude
  analyses varying PID, tracking, and reconstruction efficiencies according to their
  uncertainties.
\end{itemize}

\begin{itemize}
\item {\it Background}
\item[-](6) The MC background yields are varied within their uncertainties
  and the largest difference from the fits is taken as the uncertainty from
  the background level. The background shape $B(p_j)$ mentioned in Eq.~\ref{bkgpdf}, derived from another combination of five variables ($M_{K^-K^+}$, $M_{\pi^+\pi^0}$, $M_{K^+\pi^0}$, $M_{K^-\pi^+}$, and $M_{K^-K^+\pi^0}$) is considered and applied.
  The square root of the quadratic sum of these two uncertainties 
  is taken as the background uncertainty.
\end{itemize}

\begin{itemize}
\item {\it Fit bias}
\item[-](7) The uncertainty due to the fit procedure is evaluated by
  studying signal MC samples. An ensemble of 300 signal MC samples are generated
  according to the nominal result in this analysis. After applying the
  selection criteria, each of these samples has the same size as the
  data sample and is used to perform the same amplitude analysis. The pull of each parameter is defined as
  $\frac{\text{Out}(i)-\text{In}(i)}{\sigma_{\text{stat.}}(i)}$, where
  $i$ denotes different parameters, $\text{In}(i)$ denotes the input
  value as taken from the nominal fit to data, $\text{Out}(i)$ is the
  value obtained from the fit to a signal MC sample and
  $\sigma_{\text{stat.}}(i)$ is the
  corresponding statistical uncertainty. For each
  parameter, 300 pull values are obtained and the
  deviation of their average from zero 
  is taken as the systematic uncertainty.
\end{itemize}

\begin{table*}[htb]
 \centering
 \caption{The phase systematic uncertainty sources (in units of
   statistical standard deviations) are (1) mass and width, (2) shape
   of the $\rho^+$ meson, (3) parameters of the $a_0^0(980)$ meson,
   (4) $R$ value, (5) experimental effects, (6) background, and (7) fit
   bias.}
 \label{pulldistribution}
  \vspace{2mm} 
\renewcommand\arraystretch{1.2}
 \begin{tabular}{cccccccccc}
\hline\hline
                Phase ($\phi$)                                          &    1     &   2   &   3   &  4   &   5  &  6    &  7    &Total  \\       
  \hline                                                                                                                                           
    $D_s^+[S]\to \phi\rho^+$                                            &0 (fixed) &       &       &      &      &       &       &       \\
    $D_s^+[P]\to \phi\rho^+$                                            &0.11      & 0.05  &0.00   &0.39  & 0.00 & 0.02  & 0.06  & 0.41  \\
    $D_s^+[D]\to \phi\rho^+$                                            &0.10      & 0.23  &0.03   &0.56  & 0.06 & 0.06  & 0.06  & 0.62  \\\hline  
    $D_s^+[S]\to \bar{K}^{*0}K^{*+}$                                    &0.74      & 0.10  &0.09   &0.26  & 0.08 & 0.18  & 0.06  & 0.82  \\
    $D_s^+[P]\to \bar{K}^{*0}K^{*+}$                                    &0.49      & 0.05  &0.05   &0.20  & 0.04 & 0.08  & 0.06  & 0.55  \\
    $D_s^+[D]\to \bar{K}^{*0}K^{*+}$                                    &0.34      & 0.01  &0.01   &0.11  & 0.00 & 0.03  & 0.06  & 0.37  \\\hline  
  $D_s^+\to \bar{K}_1^0(1270)K^+$,~$\bar{K}_1^0(1270)\to K^-\rho^+$     &1.95      & 0.05  &0.13   &0.63  & 0.13 & 0.24  & 0.06  & 2.07  \\\hline  
  $D_s^+\to \bar{K}_1^0(1270)K^+$,~$\bar{K}_1^0(1270)[S]\to K^{*}\pi$   &0.92      & 0.17  &0.17   &0.47  & 0.18 & 0.23  & 0.06  & 1.10  \\
  $D_s^+\to \bar{K}_1^0(1270)K^+$,~$\bar{K}_1^0(1270)[D]\to K^{*}\pi$   &1.24      & 0.18  &0.18   &0.32  & 0.20 & 0.29  & 0.06  & 1.36  \\\hline  
  $D_s^+\to \bar{K}_1^0(1400)K^+$,~$\bar{K}_1^0(1400)[S]\to K^{*}\pi$   &0.74      & 0.01  &0.00   &0.14  & 0.00 & 0.01  & 0.06  & 0.76  \\\hline  
    $D_s^+\to a_0^0(980)\rho^+$                                         &0.69      & 0.14  &0.02   &0.77  & 0.05 & 0.05  & 0.06  & 1.05  \\\hline  
  $D_s^+\to f_1(1420)\pi^+$,~$f_1(1420)\to K^{*\mp}K^{\pm}$             &0.42      & 0.10  &0.10   &0.46  & 0.15 & 0.09  & 0.07  & 0.67  \\\hline  
  $D_s^+\to f_1(1420)\pi^+$,~$f_1(1420)\to a_0^0(980)\pi^0$             &0.30      & 0.03  &0.12   &0.31  & 0.02 & 0.11  & 0.05  & 0.47  \\\hline  
  $D_s^+\to \eta(1475)\pi^+$,~$\eta(1475)\to a_0^0(980)\pi^0$           &0.43      & 0.08  &0.06   &0.45  & 0.05 & 0.01  & 0.08  & 0.63  \\        
\hline\hline
 \end{tabular}
\end{table*}

\begin{table*}[htb]
 \centering
 \caption{The FF systematic uncertainty sources (in units of
   statistical standard deviations) are (1) mass and width, (2) shape
   of $\rho^+$ meson, (3) parameters of $a_0^0(980)$ meson, (4) $R$
   value, (5) experimental effects, (6) background, and (7) fit
   bias. The last row is the systematic uncertainty of the ratio
   $\frac{{\cal B}(D_s^+\to \bar{K}_1^0(1270)K^+,~\bar{K}_1^0(1270)\to
     K^{*-}\pi^+)}{{\cal B}(D_s^+\to
     \bar{K}_1^0(1270)K^+,~\bar{K}_1^0(1270)\to K^-\rho^+)}$.}
 \label{pulldistribution2}
  \vspace{2mm} 
\renewcommand\arraystretch{1.2}
 \begin{tabular}{cccccccccc}
\hline\hline
       FF                                                                                                                                                     &   1   &   2   &  3   &  4   &  5   &   6   &  7 &Total \\                     
  \hline                                                                                                                                                                                                                                       
   $D_s^+[S]\to \phi\rho^+$                                                                                                                                   &1.09   &0.33   &0.33  &0.54  &0.58  &0.55   &0.07&1.53     \\                     
   $D_s^+[P]\to \phi\rho^+$                                                                                                                                   &0.28   &0.08   &0.08  &0.14  &0.04  &0.11   &0.07&0.36  \\                     
   $D_s^+[D]\to \phi\rho^+$                                                                                                                                   &0.26   &0.08   &0.09  &0.12  &0.08  &0.09   &0.13&0.36  \\                     
   $D_s^+\to \phi\rho^+$                                                                                                                                      &1.46   &0.33   &0.45  &1.00  &0.88  &0.72   &0.07&2.18   \\\hline               
   $D_s^+[S]\to \bar{K}^{*0}K^{*+}$                                                                                                                           &0.19   &0.03   &0.07  &0.12  &0.23  &0.09   &0.06&0.35  \\                     
   $D_s^+[P]\to \bar{K}^{*0}K^{*+}$                                                                                                                           &0.34   &0.04   &0.09  &0.22  &0.20  &0.14   &0.07&0.48    \\                     
   $D_s^+[D]\to \bar{K}^{*0}K^{*+}$                                                                                                                           &0.13   &0.01   &0.03  &0.23  &0.11  &0.04   &0.13&0.32  \\                     
   $D_s^+\to \bar{K}^{*0}K^{*+}$                                                                                                                              &0.56   &0.12   &0.19  &0.21  &0.44  &0.28   &0.08&0.83  \\\hline               
   $D_s^+\to \bar{K}_1^0(1270)K^+$,~$\bar{K}_1^0(1270)\to K^-\rho^+$                                                                                          &0.62   &0.30   &0.18  &0.40  &0.25  &0.27   &0.08&0.90  \\\hline               
   \footnotesize{$D_s^+\to \bar{K}_1^0(1270)K^+$},~\footnotesize{$\bar{K}_1^0(1270)[S]\to \bar{K}^{*0}\pi^0$}                                                 &0.55   &0.14   &0.07  &0.31  &0.00  &0.08   &0.07&0.67  \\                     
   \footnotesize{$D_s^+\to \bar{K}_1^0(1270)K^+$},~\footnotesize{$\bar{K}_1^0(1270)[S]\to K^{*-}\pi^+$}                                                       &0.54   &0.14   &0.07  &0.30  &0.00  &0.08   &0.07&0.65  \\                     
   $D_s^+\to \bar{K}_1^0(1270)K^+$,~$\bar{K}_1^0(1270)[S]\to K^{*}\pi$                                                                                        &0.54   &0.14   &0.07  &0.29  &0.00  &0.08   &0.07&0.64  \\\cdashline{1-9}[0.8pt/2pt] 
   \footnotesize{$D_s^+\to \bar{K}_1^0(1270)K^+$},~\footnotesize{$\bar{K}_1^0(1270)[D]\to \bar{K}^{*0}\pi^0$}                                                 &0.44   &0.29   &0.11  &0.67  &0.09  &0.16   &0.05&0.88  \\                           
   \footnotesize{$D_s^+\to \bar{K}_1^0(1270)K^+$},~\footnotesize{$\bar{K}_1^0(1270)[D]\to K^{*-}\pi^+$}                                                       &0.43   &0.29   &0.11  &0.67  &0.09  &0.16   &0.05&0.87  \\                           
   $D_s^+\to \bar{K}_1^0(1270)K^+$,~$\bar{K}_1^0(1270)[D]\to K^{*}\pi$                                                                                        &0.43   &0.29   &0.11  &0.67  &0.09  &0.16   &0.05&0.87  \\\cdashline{1-9}[0.8pt/2pt] 
   $D_s^+\to \bar{K}_1^0(1270)K^+$,~$\bar{K}_1^0(1270)\to K^{*}\pi$                                                                                           &0.71   &0.32   &0.16  &0.71  &0.08  &0.21   &0.10&1.10  \\\hline                     
   \footnotesize{$D_s^+\to \bar{K}_1^0(1400)K^+$},~\footnotesize{$\bar{K}_1^0(1400)[S]\to \bar{K}^{*0}\pi^0$}                                                 &0.76   &0.21   &0.27  &0.49  &0.36  &0.44   &0.09&1.12   \\                           
   \footnotesize{$D_s^+\to \bar{K}_1^0(1400)K^+$},~\footnotesize{$\bar{K}_1^0(1400)[S]\to K^{*-}\pi^+$}                                                       &0.77   &0.21   &0.27  &0.49  &0.36  &0.44   &0.09&1.13   \\                           
   $D_s^+\to \bar{K}_1^0(1400)K^+$,~$\bar{K}_1^0(1400)[S]\to K^{*}\pi$                                                                                        &0.76   &0.21   &0.27  &0.49  &0.36  &0.44   &0.09&1.12   \\\hline                     
   $D_s^+\to a_0^0(980)\rho^+$                                                                                                                                &0.74   &0.28   &0.27  &0.30  &0.29  &0.47   &0.06&1.05   \\\hline                     
   \footnotesize{$D_s^+\to f_1(1420)\pi^+$},~\footnotesize{$f_1(1420)\to K^{*-}K^+$}                                                                          &0.39   &0.10   &0.11  &0.37  &0.19  &0.18   &0.11&0.62  \\                           
   \footnotesize{$D_s^+\to f_1(1420)\pi^+$},~\footnotesize{$f_1(1420)\to K^{*+}K^-$}                                                                          &0.37   &0.10   &0.11  &0.37  &0.19  &0.18   &0.11&0.61  \\                           
   $D_s^+\to f_1(1420)\pi^+$,~$f_1(1420)\to K^{*\mp}K^{\pm}$                                                                                                  &0.38   &0.10   &0.11  &0.33  &0.19  &0.18   &0.11&0.59   \\\hline                     
   $D_s^+\to f_1(1420)\pi^+$,~$f_1(1420)\to a_0^0(980)\pi^0$                                                                                                  &0.22   &0.07   &0.08  &0.16  &0.13  &0.07   &0.07&0.33   \\\hline                     
   $D_s^+\to \eta(1475)\pi^+$,~$\eta(1475)\to a_0^0(980)\pi^0$                                                                                                &0.72   &0.31   &0.29  &0.40  &0.33  &0.39   &0.10&1.07   \\\hline                     
   $\frac{{\cal B}(D_s^+\to \bar{K}_1^0(1270)K^+,~\bar{K}_1^0(1270)\to K^{*-}\pi^+)}{{\cal B}(D_s^+\to \bar{K}_1^0(1270)K^+,~\bar{K}_1^0(1270)\to K^-\rho^+)}$&0.75   &0.39   &0.19  &0.65  &0.14  &0.25   &0.08&1.13  \\     %
\hline\hline
 \end{tabular}
\end{table*}

\section{\boldmath Branching fraction
measurement}\label{branchfractionsssss} To determine the absolute BF
of the decay $D_s^+ \to K^- K^+ \pi^+ \pi^0$, the ST candidates with eight tag modes, 
as shown in Table~\ref{tagwindow}, are reconstructed and studied. Then the DT candidates are obtained by fully
reconstructing the tag channels and the signal channel.

The ST yields for each tag mode are given by
\begin{eqnarray}
\begin{aligned}
\label{eq:ST_yield} 
N_{\text{ST}} = 2N_{D_s^+D_s^-}{\cal B}_{\text{tag}}\varepsilon_{\text{ST}}\,,
\end{aligned}
\end{eqnarray}
and the DT yields are given by
\begin{eqnarray}
\begin{aligned}
\label{eq:DT_yield}
N_{\text{DT}} = 2N_{D_s^+D_s^-}{\cal B}_{\text{tag}}{\cal B}_{\text{sig}}\varepsilon_{\text{DT}}\,, 
\end{aligned}
\end{eqnarray}
where $N_{D_s^+D_s^-}$ is the total number of $D_s^+D_s^-$ pairs produced,
${\cal B}_{\text{tag(sig)}}$ is the BF of the tag (signal) side,
and $\varepsilon_{\text{DT(ST)}}$ is the DT (ST) efficiency.

The BF of the signal side is determined by
\begin{eqnarray}
\begin{aligned}
\label{eq:DTag_BF}
{\cal B}_{\text{sig}} = \frac{N_{\text{DT}}}{{\cal B}_{\pi^0\to \gamma\gamma}\sum\limits_{i}N^{i}_{\text{ST}}\varepsilon^{i}_{\text{DT}}/\varepsilon^{i}_{\text{ST}}} \,,
\end{aligned}
\end{eqnarray}
where the $N_{\text{DT}}$ and $N^{i}_{\text{ST}}$ yields are obtained
from the data sample, while $\varepsilon^{i}_{\text{DT}}$ and
$\varepsilon^{i}_{\text{ST}}$ are obtained from the generic MC samples, where
$i$ indicates the tag mode. In particular,
$\varepsilon^{i}_{\text{DT}}$ is determined by the amplitude analysis
model used in the generic MC samples.

The signal BF ${\cal B}_{\text{sig}}$ is determined by
\begin{eqnarray}
\begin{aligned} \label{tot:DTag_BF} {\cal B}_{\text{sig}} =
\frac{\sum\limits_n N_{n\text{DT}}}{{\cal B}_{\pi^0\to
\gamma\gamma}\sum\limits_{n}\sum\limits_{i}N^{i}_{n\text{ST}}\varepsilon^{i}_{n\text{DT}}/\varepsilon^{i}_{n\text{ST}}}
\,,
\end{aligned}
\end{eqnarray}
where $i$ denotes the tag mode and
$n$ indicates the data sample at 4.178 GeV, 4.189-4.219
GeV or 4.226 GeV. For the numerator, $\sum\limits_n N_{n\text{DT}}$, 
the combined data sample is fitted to obtain the total DT data yield.

\subsection{\boldmath Event Selection}
For the BF measurement, it is
necessary to guarantee that the DT sample is a strict subset of the ST
sample. Therefore, the ST candidates are selected ahead of the selection of DT
candidates. For this measurement, the event selection criteria are
relaxed or modified in order to increase the signal yield. Here, the
background level does not play as crucial a role as in the amplitude
analysis.

In order to reject the soft pions from $D^*$ decays, all the $\pi$ mesons
are required to satisfy $P_{\pi}> 100$~MeV/$c$, and the $\chi^2$
of the kinematic fit for the $\pi^0\to\gamma\gamma$ decay must be less
than 20. The new criteria for selecting $K_S^0$ are $487 <
M_{\pi^+\pi^-} < 511~({\rm MeV}/c^2$) and that the vertex fit $\chi^2$
must be less than 100.

For the ST selection, if there are multiple candidates for a tag mode, 
the one with $M_{\rm rec}$ closest to the nominal
$M_{D_s^{*\pm}}$~\cite{PDG} is retained. The $M_{\rm rec}$ windows are given in
Table~\ref{mrec}. If the $D_s^+$ meson and the $D_s^-$ meson can be
simultaneously reconstructed as ST in an event, both of them are
accepted. After the ST selection, if multiple signal candidates are
obtained, the one with average mass $\bar{M}$ ($=
(M_{D_s^+}+M_{D_s^-})/2$) closest to the nominal $M_{D_s^{\pm}}$ is
chosen. $M_{D_s^{\pm}}$ of every candidate must lie in the interval [1.87,~2.06]
GeV/$c^2$, and events with both $M_{\rm rec}$ for the $D_s^+$ meson
and $M_{\rm rec}$ for the $D_s^-$ meson smaller than 2.1 GeV are
rejected.

\subsection{\boldmath Data Yields, Efficiencies and BFs}
\begin{figure*}[htbp]  
    \mbox{
    \begin{overpic}[width=5.7cm,height=4.6cm,angle=0]{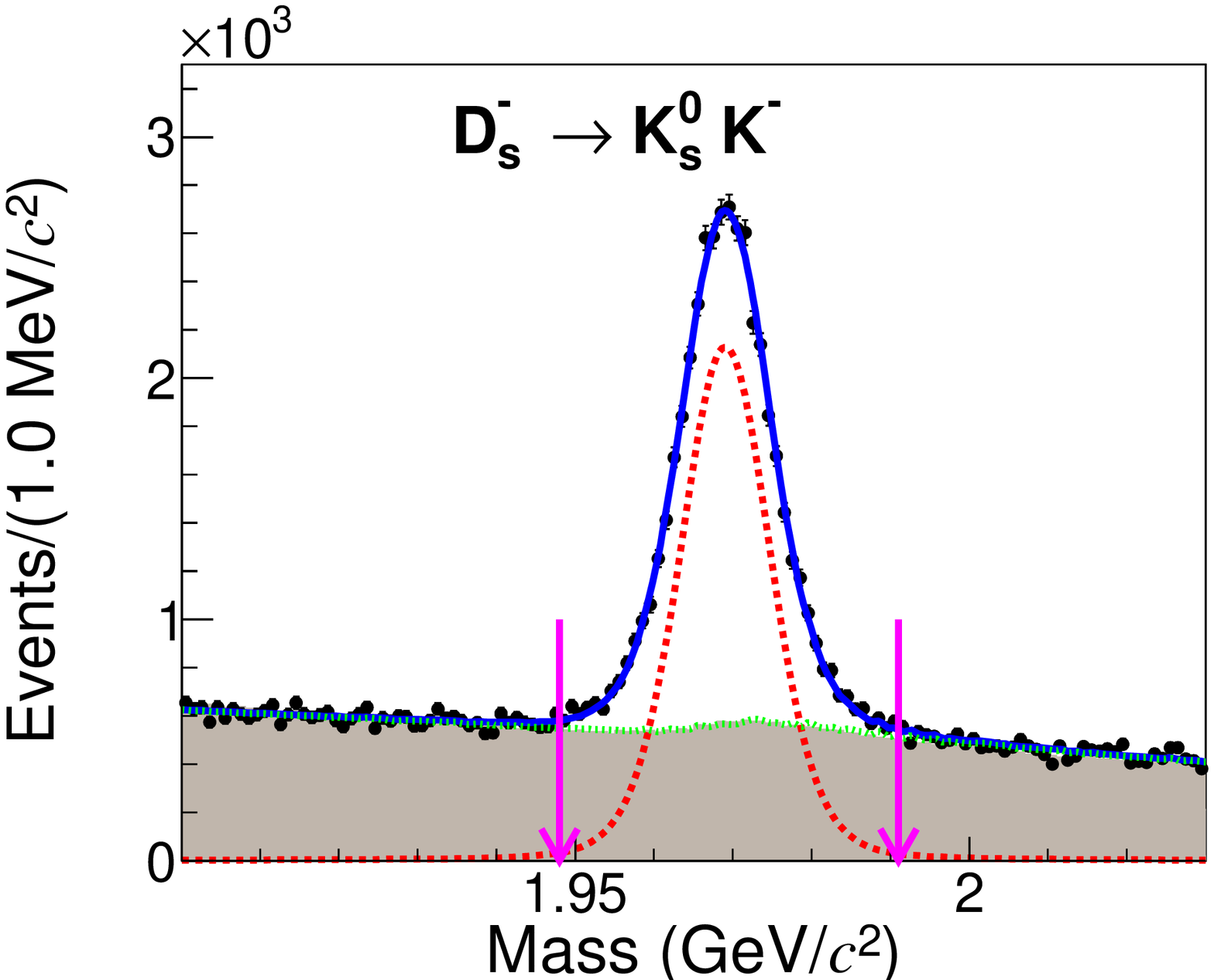}
    \end{overpic}
    \begin{overpic}[width=5.7cm,height=4.6cm,angle=0]{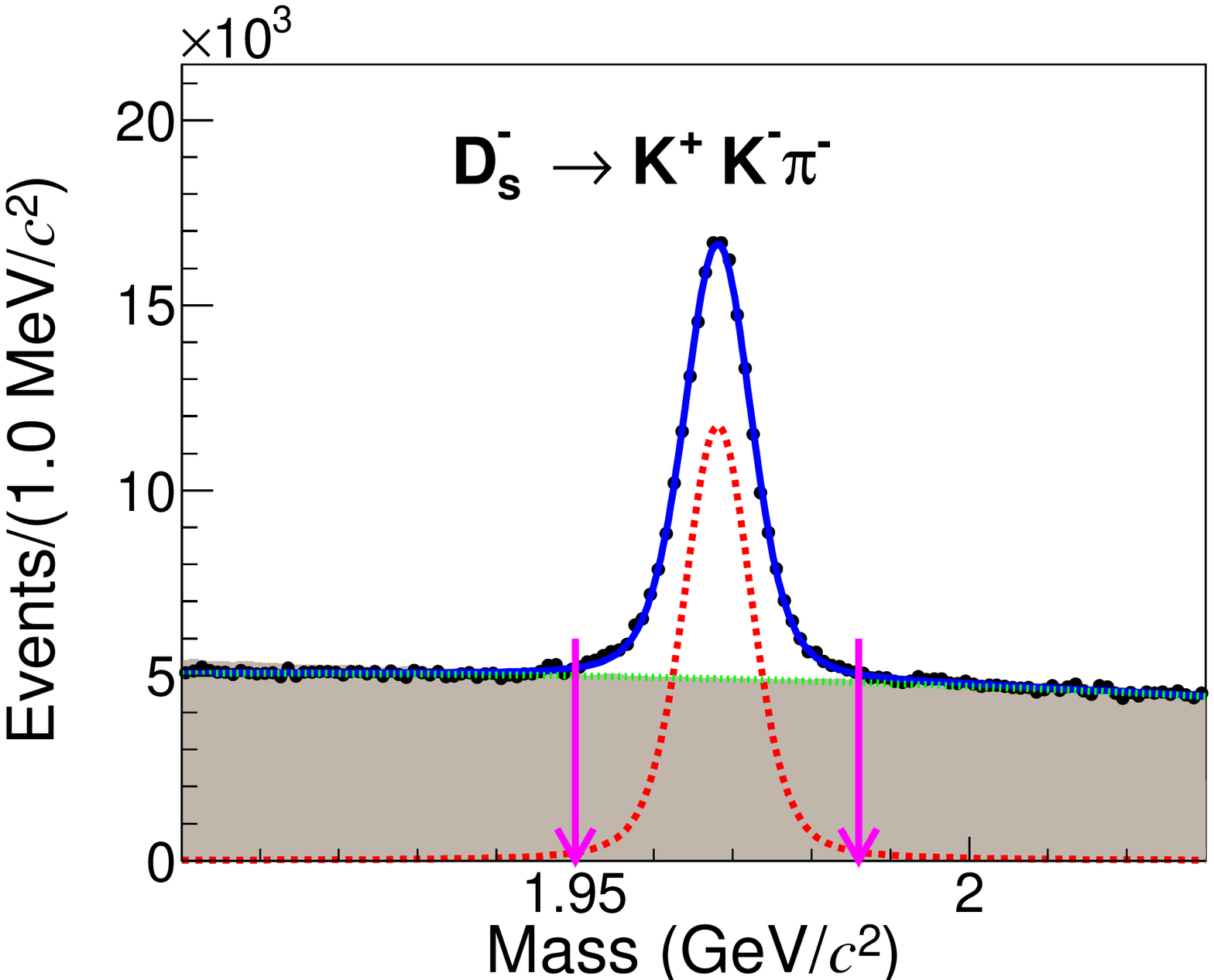}
    \end{overpic}
    \begin{overpic}[width=5.7cm,height=4.6cm,angle=0]{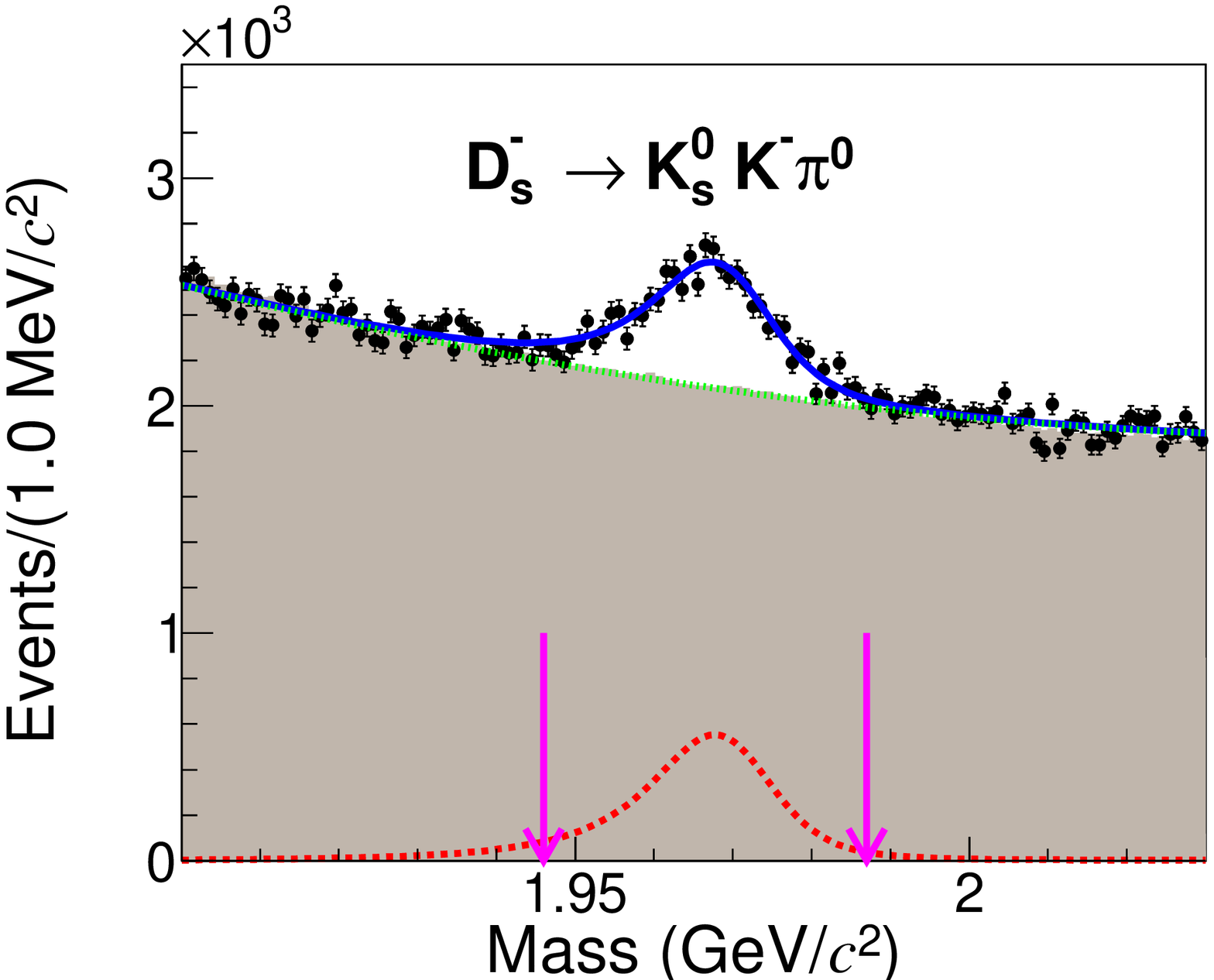}
    \end{overpic} }
    \mbox{
    \begin{overpic}[width=5.7cm,height=4.6cm,angle=0]{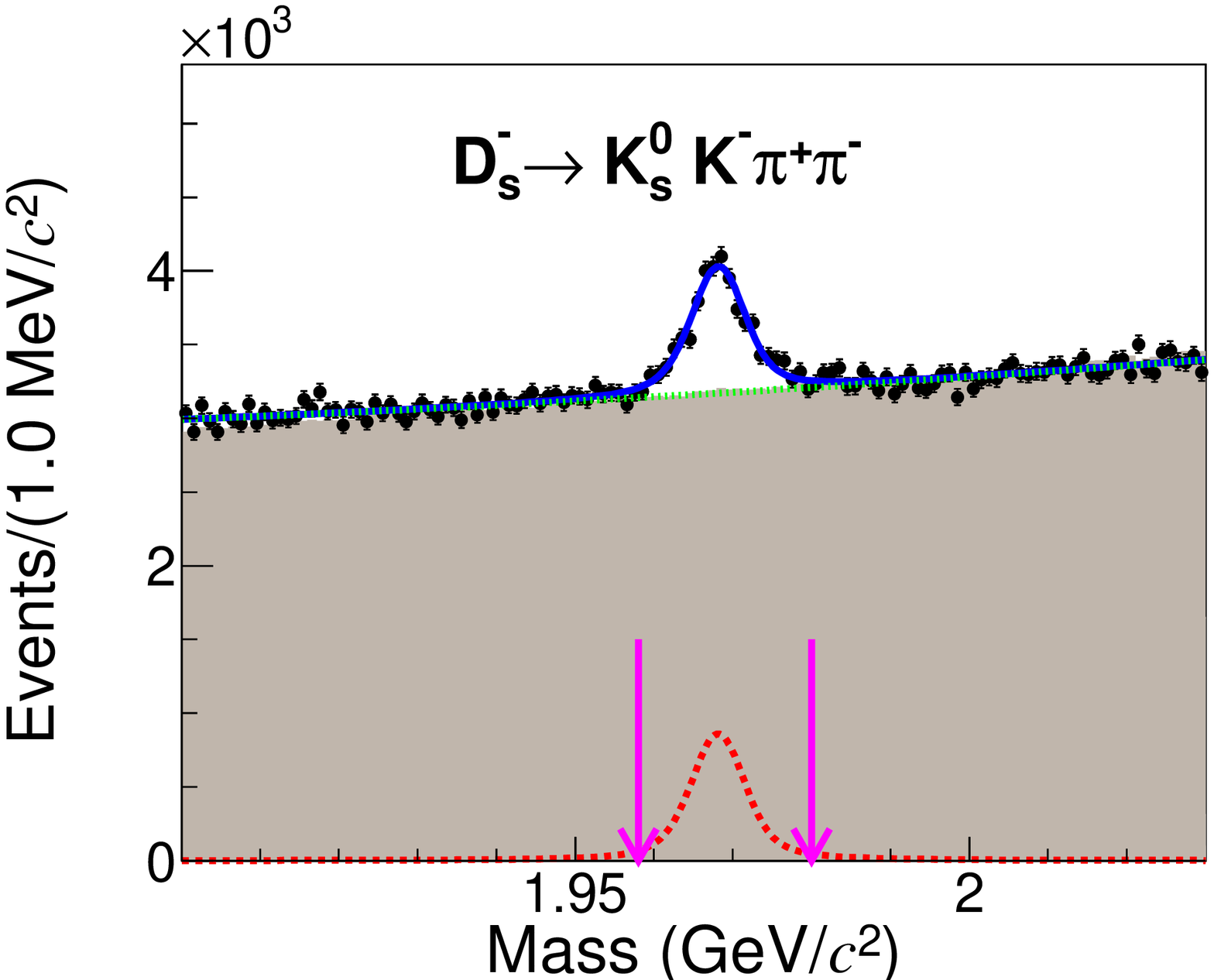}
    \end{overpic}
    \begin{overpic}[width=5.7cm,height=4.6cm,angle=0]{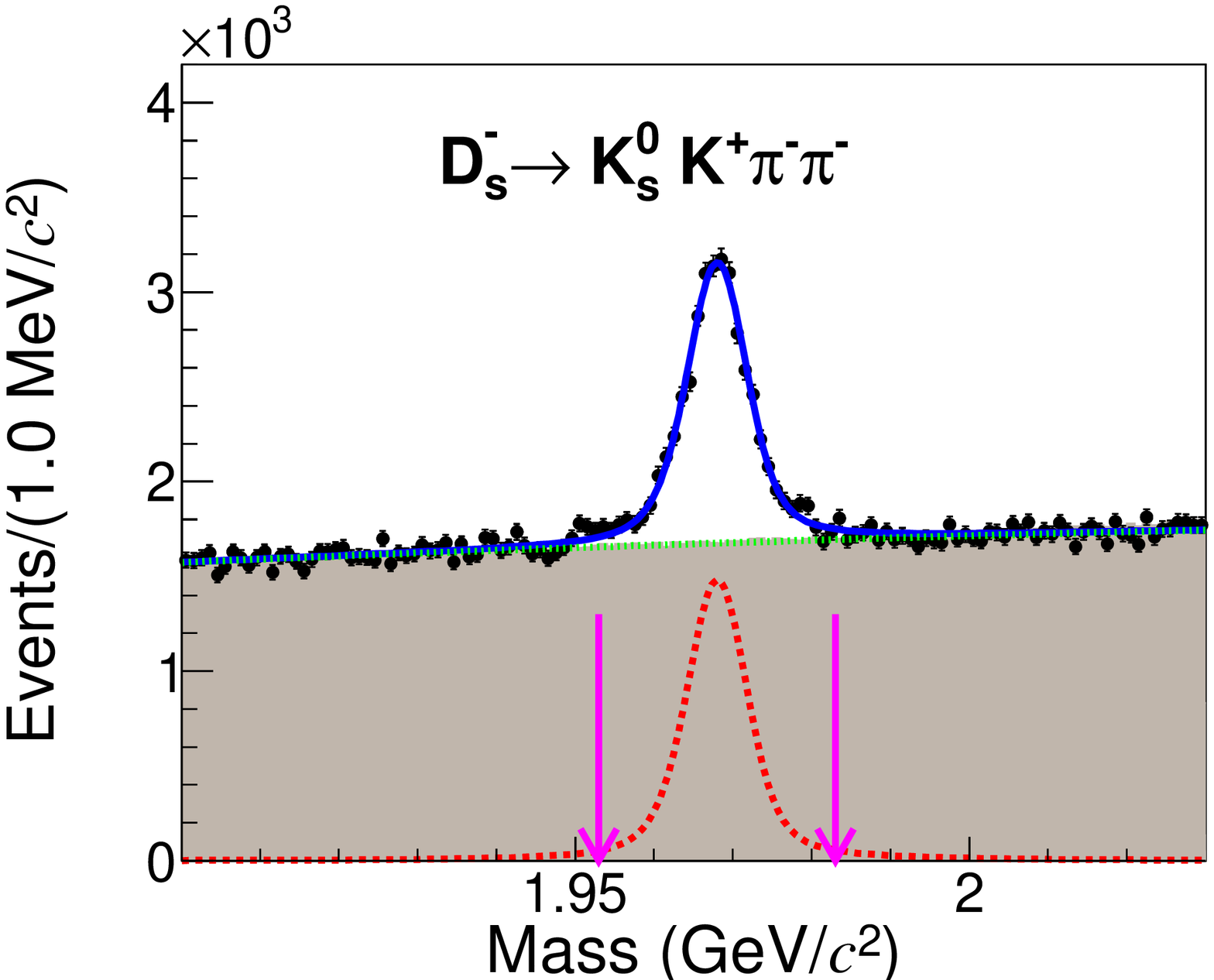}
    \end{overpic}
    \begin{overpic}[width=5.7cm,height=4.6cm,angle=0]{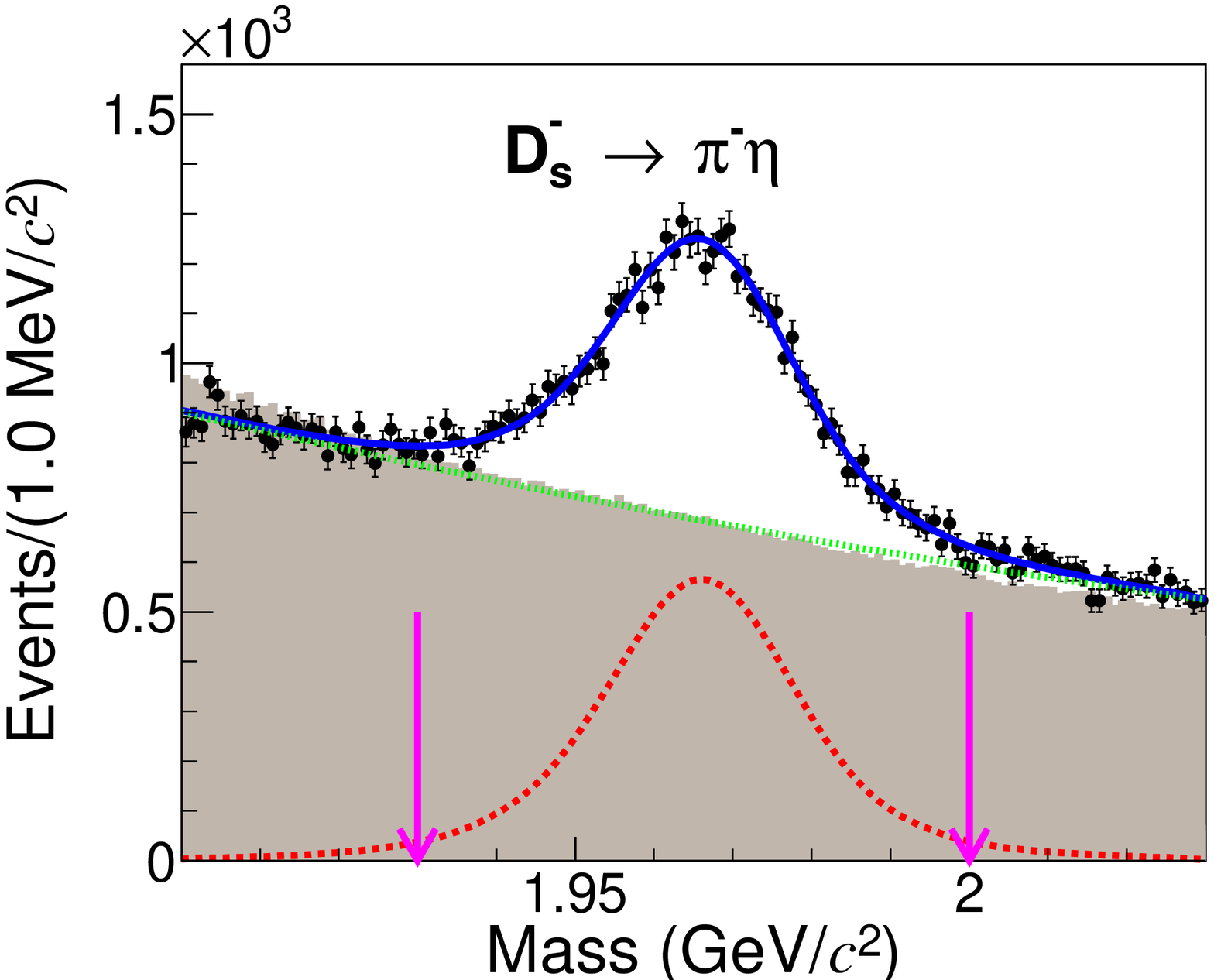}
    \end{overpic} }
    \mbox{
   \hskip -5.9cm{ \begin{overpic}[width=5.7cm,height=4.6cm,angle=0]{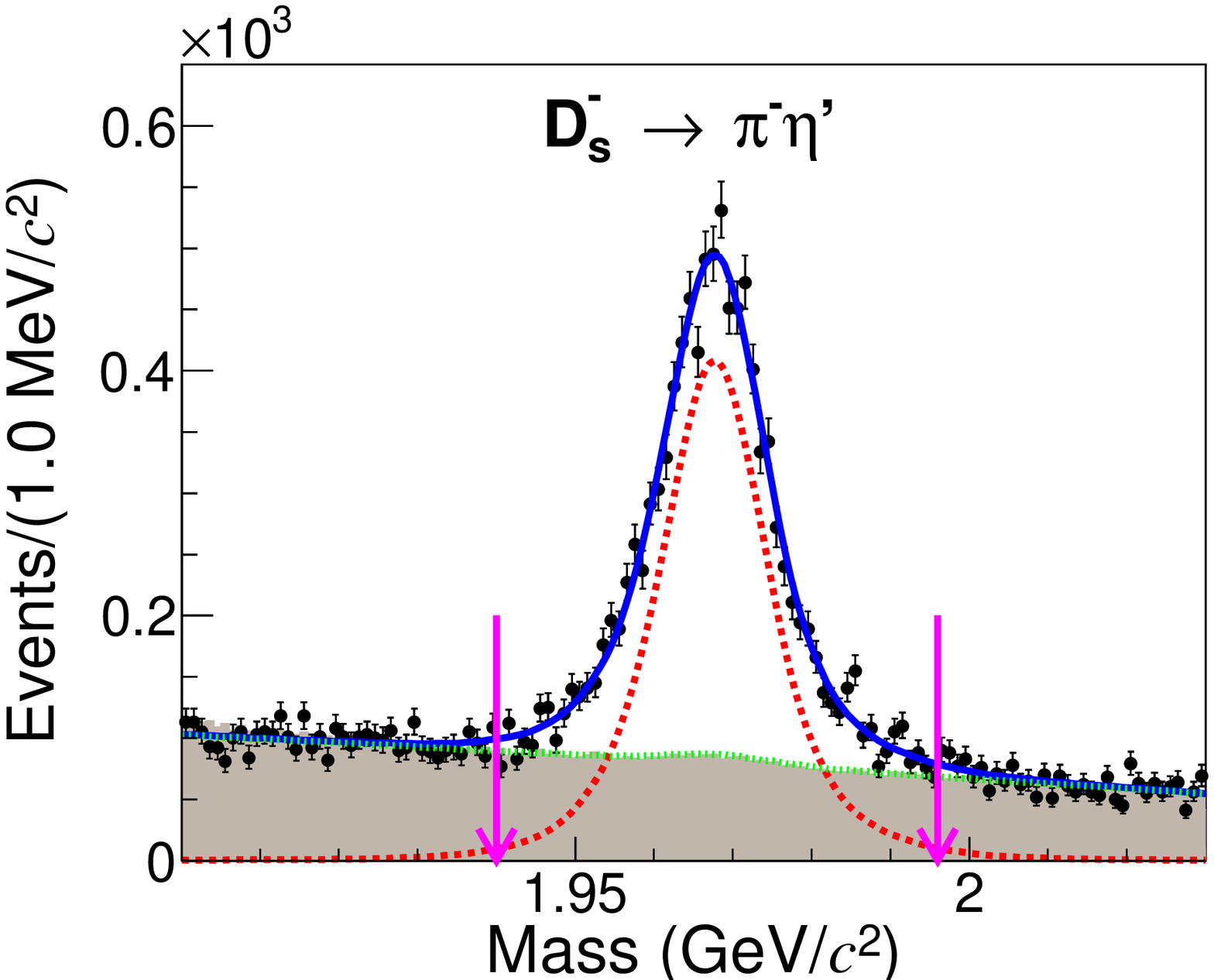}
    \end{overpic}
    \begin{overpic}[width=5.7cm,height=4.6cm,angle=0]{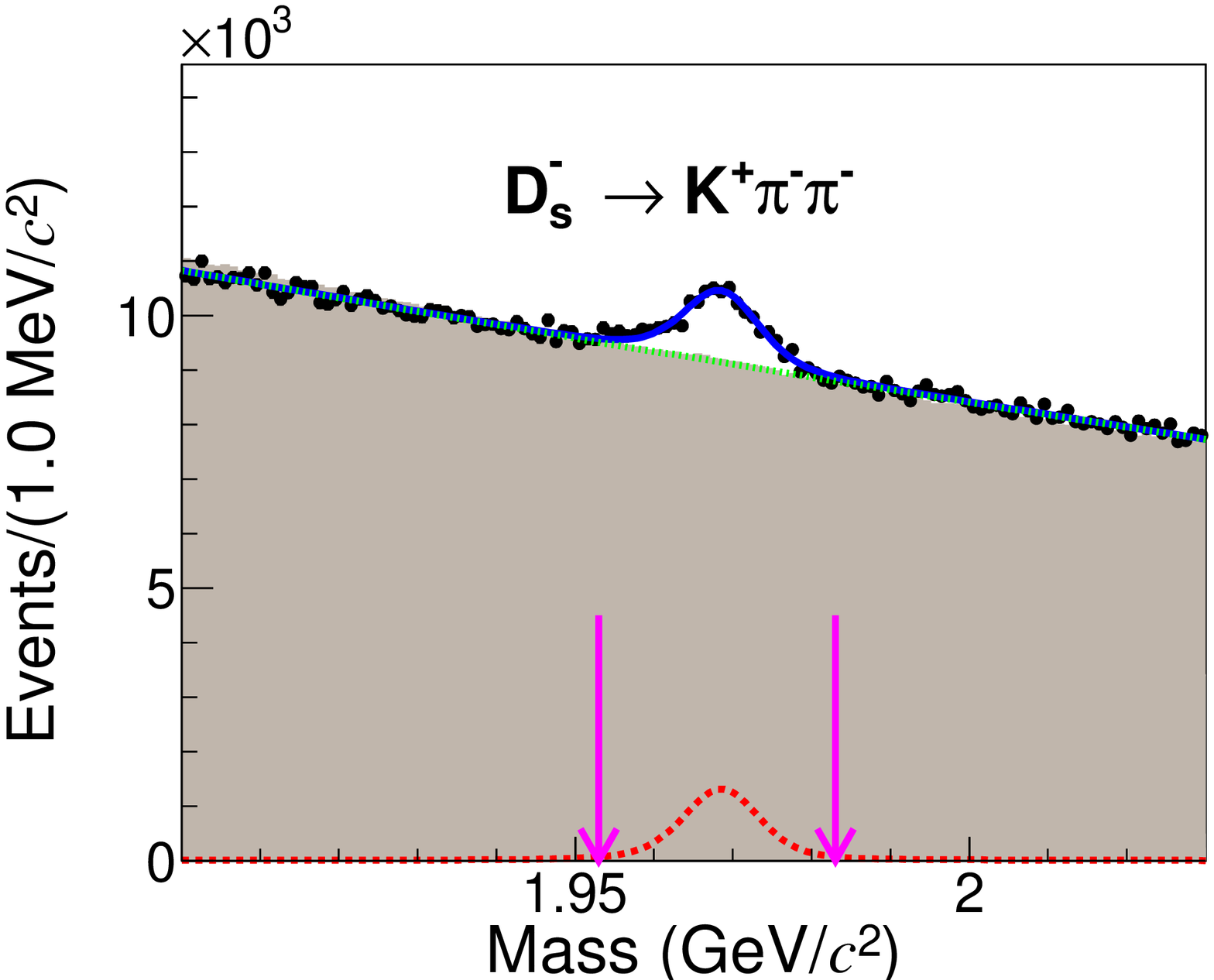}
    \end{overpic}} } 
\caption{Fits to the $M_{D_s^-}$ distributions of ST candidates
  selected from the 4.178 GeV data sample, where the dots with error
  bars are data, the solid blue curve shows the best fit, the red
  dotted curve shows the signal shape, the green dashed line shows the
  shape of the combinatorial backgrounds, the brown area shows the
  background estimated by the generic MC samples, and the pairs of pink
  arrows are the mass windows. In the plots for 
  $D_s^-\to K_S^0K^-$ and $D_s^-\to \pi^-\eta^{\prime}$
  decays, the green dashed lines include contributions from
  $D^-\to K_S^0\pi^-$
  and $D_s^-\to \eta\pi^+\pi^-\pi^-$ backgrounds, respectively.}
\label{singletagfig}
\end{figure*}

\begin{figure}[htbp]
\centering
    \mbox{
    \begin{overpic}[width=7.0cm,height=5.6cm,angle=0]{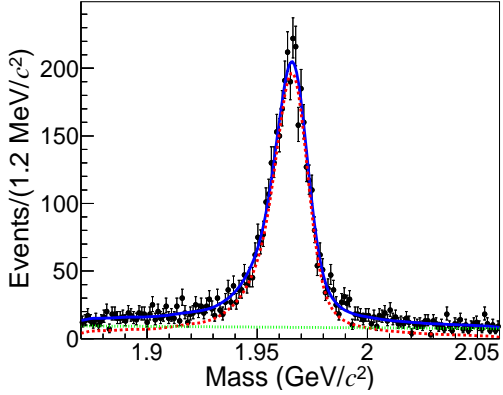}
    \end{overpic}
    }
\caption{Invariant mass distribution of the DT $D_s^+ \to
  K^-K^+\pi^+\pi^0$ events. The black dots with error bars are
  data. The red dashed line represents the MC-simulated shape
  convolved with a Gaussian function. The green dashed line represents
  the MC background shape, which is fitted by a $1^{\rm st}$-order
  Chebychev polynomial. The blue solid line represents the total
  fitted shape.}
\label{DTyieldsaa}
\end{figure}

The ST yields are determined from fits to the $M_{D_s^-}$
distributions of data, as shown in Fig.~\ref{singletagfig}. In the
fits, an MC-simulated shape convolved with a Gaussian function is used
to describe the signal shape of $M_{D_s^-}$ and a $2^{\rm nd}$-order
polynomial function to describe the combinatorial background. For the
tag mode $D_s^-\to K^0_S K^-$, there is some peaking background coming
from $D^-\to K_S^0\pi^-$. The shape of this background is taken from 
the generic MC samples and added to the fit, leaving its yield floating. For
the tag mode $D_s^-\to \pi^- \eta^{\prime}$, there is peaking
background coming from $D_s^-\to \eta\pi^+\pi^-\pi^-$. The
shape and the yield of this background are taken from the generic MC samples and added to the fit.
The DT yields are obtained from an unbinned fit to the
signal $D_s^+$ mass spectrum of the combined data sample,
which is shown in Fig.~\ref{DTyieldsaa}. The number of $D_s^+\to K^-K^+\pi^+\pi^0$ decays is determined to be
$\sum\limits_n N_{n\text{DT}}= 4365\pm 83$. Tables~\ref{effiandyield}-\ref{effiandyield2} summarize the ST
efficiencies, DT efficiencies, and ST yields in data samples at
4.178-4.226 GeV.

\begin{table*}[htb]
 \centering
 \caption{The efficiencies and ST yields at $E_{\rm cm}$ = 4.178 GeV.}
 \label{effiandyield}
\renewcommand\arraystretch{1.2}
 \begin{tabular}{ccccc}
\hline\hline
  Tag mode                                                        & Mass window (GeV/$c^2$) & $~N_{\rm ST}$ & $\varepsilon_{\text{ST}} (\%)$ & $\varepsilon_{\text{DT}} (\%)$ \\  
  \hline
  $D_s^-\to K_S^0 K^-$                                            &  [1.948,~1.991]  &$\,~31668\pm315$    &$46.95\pm0.07$    &$8.75\pm0.09$  \\
  $D_s^-\to K^+K^-\pi^-$                                          &  [1.950,~1.986]  &$135867\pm610$    &$39.00\pm0.03$    &$7.09\pm0.03$  \\
  $D_s^-\to K_S^0K^-\pi^0$                                        &  [1.946,~1.987]  &$\,~11284\pm512$    &$15.32\pm0.11$    &$2.92\pm0.05$  \\
  $D_s^-\to K_S^0K^-\pi^+\pi^-$                                   &  [1.958,~1.980]  &$\,\;\;\;8032\pm273$    &$20.29\pm0.12$    &$3.36\pm0.07$  \\
  $D_s^-\to K_S^0K^+\pi^-\pi^-$                                   &  [1.953,~1.983]  &$\,~15645\pm289$    &$21.70\pm0.06$    &$3.76\pm0.05$  \\
  $D_s^-\to \pi^-\eta_{\gamma\gamma}$                             &  [1.930,~2.000]  &$\,~18071\pm560$    &$43.07\pm0.15$    &$7.92\pm0.10$  \\
  $D_s^-\to \pi^-\eta^{\prime}_{\pi^+\pi^-\eta_{\gamma\gamma}}$   &  [1.940,~1.996]  &$\,\;\;\;7629\pm147$    &$18.72\pm0.06$    &$3.19\pm0.06$  \\
  $D_s^-\to K^-\pi^+\pi^-$                                        &  [1.953,~1.983]  &$\,~16942\pm548$    &$45.80\pm0.22$    &$8.39\pm0.10$  \\
\hline\hline
 \end{tabular}
\end{table*}

\begin{table*}[htb]
 \centering
 \caption{The efficiencies and ST yields at $E_{\rm cm}$ = 4.189-4.219 GeV.}
 \label{effiandyield1}
\renewcommand\arraystretch{1.2}
 \begin{tabular}{ccccc}
\hline\hline
  Tag mode                                                       & Mass window (GeV/$c^2$) & $~N_{\rm ST}$ & $\varepsilon_{\text{ST}} (\%)$ & $\varepsilon_{\text{DT}} (\%)$ \\  
  \hline
  $D_s^-\to K_S^0 K^-$                                           &  [1.948,~1.991]  &$18304\pm260$    &$46.87\pm0.09$   &$9.08\pm0.11$  \\
  $D_s^-\to K^+K^-\pi^-$                                         &  [1.950,~1.986]  &$80417\pm508$    &$38.82\pm0.04$   &$7.28\pm0.04$  \\
  $D_s^-\to K_S^0K^-\pi^0$                                       &  [1.946,~1.987]  &$\,~6730\pm462$    &$14.88\pm0.15$   &$3.11\pm0.07$  \\
  $D_s^-\to K_S^0K^-\pi^+\pi^-$                                  &  [1.958,~1.980]  &$\,~5252\pm285$    &$20.07\pm0.16$   &$3.32\pm0.08$  \\
  $D_s^-\to K_S^0K^+\pi^-\pi^-$                                  &  [1.953,~1.983]  &$\,~8923\pm230$    &$21.53\pm0.08$   &$3.86\pm0.07$  \\
  $D_s^-\to \pi^-\eta_{\gamma\gamma}$                            &  [1.930,~2.000]  &$10034\pm355$    &$42.37\pm0.21$   &$8.15\pm0.13$  \\
  $D_s^-\to \pi^-\eta^{\prime}_{\pi^+\pi^-\eta_{\gamma\gamma}}$  &  [1.940,~1.996]  &$\,~4382\pm112$    &$18.66\pm0.07$   &$3.45\pm0.09$  \\
  $D_s^-\to K^-\pi^+\pi^-$                                       &  [1.953,~1.983]  &$10051\pm529$    &$45.38\pm0.30$   &$8.41\pm0.13$  \\
\hline\hline
 \end{tabular}
\end{table*}

\begin{table*}[htb]
 \centering
 \caption{The efficiencies and ST yields at $E_{\rm cm}$ = 4.226 GeV.}
 \label{effiandyield2}
\renewcommand\arraystretch{1.2}
 \begin{tabular}{ccccc}
\hline\hline 
  Tag mode                                                       & Mass window (GeV/$c^2$) & $~N_{\rm ST}$ & $\varepsilon_{\text{ST}} (\%)$ & $\varepsilon_{\text{DT}} (\%)$ \\  
  \hline
  $D_s^-\to K_S^0 K^-$                                           &  [1.948,~1.991]  &$\,~6550\pm159$    &$46.42\pm0.18$   &$8.81\pm0.18$  \\
  $D_s^-\to K^+K^-\pi^-$                                         &  [1.950,~1.986]  &$28290\pm328$    &$38.27\pm0.07$   &$7.30\pm0.07$  \\
  $D_s^-\to K_S^0K^-\pi^0$                                       &  [1.946,~1.987]  &$\,~2145\pm219$    &$15.22\pm0.28$   &$2.97\pm0.11$  \\
  $D_s^-\to K_S^0K^-\pi^+\pi^-$                                  &  [1.958,~1.980]  &$\,~1708\pm217$    &$19.45\pm0.30$   &$3.38\pm0.14$  \\
  $D_s^-\to K_S^0K^+\pi^-\pi^-$                                  &  [1.953,~1.983]  &$\,~3242\pm170$    &$21.31\pm0.15$   &$3.90\pm0.12$  \\
  $D_s^-\to \pi^-\eta_{\gamma\gamma}$                            &  [1.930,~2.000]  &$\,~3699\pm244$    &$41.94\pm0.40$   &$8.12\pm0.22$  \\
  $D_s^-\to \pi^-\eta^{\prime}_{\pi^+\pi^-\eta_{\gamma\gamma}}$  &  [1.940,~1.996]  &$~1646\pm75~$    &$18.45\pm0.13$   &$3.37\pm0.14$  \\
  $D_s^-\to K^-\pi^+\pi^-$                                       &  [1.953,~1.983]  &$\,~4915\pm423$    &$44.75\pm0.57$   &$8.41\pm0.22$  \\
\hline\hline
 \end{tabular}
\end{table*}

Inserting the values of the ST and DT data yields and the ST and DT
efficiencies into Eq.~\ref{tot:DTag_BF}, the BF of the
$D_s^+\to K^-K^+\pi^+\pi^0$ decay is measured to be
\begin{eqnarray}
\begin{aligned}
{\cal B}_{\text{sig}} = (5.42\pm0.10_{\rm stat.})\% \,.
\end{aligned}
\end{eqnarray}

\subsection{\boldmath Systematic Uncertainties in the BF}\label{bfzalame}
The sources of the systematic uncertainties in the BF measurement are
considered as follows.
\begin{itemize}
    \item {\it $K^{\pm}$ meson and $\pi^{\pm}$ meson tracking/PID
      efficiencies}\\ The ratios between data and MC efficiencies are
      weighted by the corresponding momentum spectra of signal MC
      events. The systematic uncertainties associated with tracking 
      efficiency and PID for each charged particle are both estimated to be 0.5\%.
      The samples used to estimate the uncertainties are mentioned in Section~\ref{PWA_systematics}.
    \item {\it $\pi^0$ meson reconstruction efficiency}\\According to the studies in Ref.~\cite{work3}, 
      this systematic uncertainty about the $\pi^0$
      reconstruction is assigned to be 2.0\%.
    \item {\it The numbers of ST $D_s^-$ candidates} \\ The BF
      measurement is not sensitive to systematic uncertainties coming
      from modifying the polynomial function order, the fit ranges or
      the bin sizes. An uncertainty of $0.56\%$ was estimated 
      from alternative fits with different signal shapes.
      According to
      Tables~\ref{effiandyield}-\ref{effiandyield2}, the total ST
      yield of the eight tag modes is $441684 \pm 1766$, corresponding
      to the relative statistical uncertainty of 0.40$\%$. The sum of
      these terms in quadrature is 0.69$\%$.
    \item {\it MC statistics}\\ The uncertainties of the ST and DT
      efficiencies are considered, but the DT uncertainties
      dominate. The uncertainty of the MC statistics is given
      by
      $\sqrt{\sum\limits_if_i(\frac{\delta\epsilon_i}{\epsilon_i})^2}$,
      where $f_i$ is the tag yield fraction and $\epsilon_i$ is the
      average DT efficiency of tag mode $i$. The related uncertainty is determined to be 0.34\%.
    \item {\it The shape of the signal $D_s^+$ mass}\\ The systematic
      uncertainty due to the shape of the signal is studied by
      fitting without the convolved Gaussian function. The difference of
      the DT yield is taken as the systematic uncertainty and is
      determined to be 0.5\%.
    \item {\it Background shape of the signal $D_s^+$ meson}\\ For the
      background shape of the signal $D_s^+$, the MC-simulated
      shape is used to replace the nominal one, and an uncertainty of 0.75\%
      is obtained.
    \item {\it Bias of the measurement method}\\ Ten updated inclusive
    generic MC samples are used as fake data to estimate the possible fit
    bias. The BF for each sample is determined, and the relative
    difference between the average of BFs and the MC truth value is
    0.16\%, which is considered negligible.

    \item {\it Amplitude model}\\ 
      The parameters (magnitudes
      and phases) of the amplitude model are randomly perturbed 400 times within their
      statistical uncertainties according to the covariant matrix of
      the nominal fit to obtain the DT efficiency distribution. Then, 
      the DT efficiency distribution is fitted with a Gaussian function. The fitted width divided by the fitted mean is 0.4\% and assigned as the systematic uncertainty arising from the amplitude model.

\end{itemize}
The systematic uncertainties in the BF are summarized in
Table~\ref{su_br}.
\begin{table}[htb]
 \centering
 \caption{The systematic uncertainties for the branching fraction measurement.}
 \label{su_br}
\renewcommand\arraystretch{1.2}
 \begin{tabular}{cc}
\hline\hline
 Source  &  Uncertainty (\%)  \\   \hline
 Tracking efficiency                      & 1.5 \\
 PID efficiency                           & 1.5 \\
 $\pi^0$ reconstruction efficiency        & 2.0 \\
 Number of $D_s^-$                        & 0.7 \\
 MC statistics                            & 0.3 \\
 Signal shape                             & 0.5 \\
 Background shape                         & 0.8 \\
 Amplitude model                          & 0.4 \\
 Total                                    & 3.2 \\
\hline\hline
 \end{tabular}
\end{table}
The total systematic uncertainty is obtained by
adding them in quadrature. Finally, the BF of the $D_s^+\to
K^-K^+\pi^+\pi^0$ decay is measured to be
\begin{eqnarray}
\begin{aligned}
\label{br_w_errnn}
{\cal B}_{\text{sig}} = (5.42\pm0.10_{\rm stat.}\pm0.17_{\rm syst.})\% . 
\end{aligned}
\end{eqnarray}

\section{\boldmath Conclusion}
This paper presents the first amplitude analysis of the decay
$D_s^+\to K^-K^+\pi^+\pi^0$. The BF $\mathcal{B}(D_s^+\to
K^-K^+\pi^+\pi^0)$ is measured to be $(5.42\pm0.10_{\rm stat.}\pm0.17_{\rm
  syst.})$\%. Using the FFs listed in Table~\ref{fitresult-table} and
Table~\ref{pulldistribution2}, the BFs for the intermediate processes
are calculated and listed in Table~\ref{summary}. The $D_s^+\to \phi
\rho^+$ and $D_s^+\to \bar{K}^{*0} K^{*+}$ decays are found to be
dominant, and the decays involving $K_1(1270), K_1(1400), \eta(1475),
f_1(1420)$, and $a_0^0(980)$ mesons are also observed with significances
larger than $4\sigma$. Compared to the PDG~\cite{PDG} values of
$\mathcal{B}(D_s^+\to K^-K^+\pi^+\pi^0)=(6.3\pm0.6)$\%,
$\mathcal{B}(D_s^+\to \phi\rho^+)=(8.4^{+1.9}_{-2.3})$\%, and
$\mathcal{B}(D_s^+\to \bar{K}^{*0}K^{*+})=(7.2\pm2.6)$\%, 
the absolute BFs
$(5.42\pm0.10_{\rm stat.}\pm0.17_{\rm syst.})$\%, $(5.59\pm0.15_{\rm
  stat.}\pm0.30_{\rm syst.})$\% and $(5.64\pm0.23_{\rm
  stat.}\pm0.27_{\rm syst.})$\% measured in this work have a much
better precision. The measurement of $\mathcal{B}(D_s^+\to
\phi\rho^+)$ is consistent with the theory
prediction~\cite{ceshijiu3} (5.70\%), while the measured BF of $D_s^+\to \bar{K}^{*0} K^{*+}$ 
decay is still much larger than its prediction (1.5\%).
The ratio $R_{K_1(1270)}=\frac{{\cal
    B}(K_1^0(1270)\to K^{*}\pi)}{{\cal B}(K_1^0(1270)\to K\rho)}$
mentioned in Table~\ref{etaao} is determined to be $0.99\pm0.15_{\rm
  stat.}\pm0.18_{\rm syst.}$ in this analysis. 
Our result is consistent with the results measured by LHCb~\cite{jhepssss} and CLEO~\cite{2012D0kkpipi1}.

\begin{table}[htp]
  \caption{The BFs of intermediate processes with final states
    $K^-K^+\pi^+\pi^0$. $K^{*}\pi$ indicates $\bar{K}^{*0}\pi^0$ and
    $K^{*-}\pi^+$. For decays with $a_0^0(980)$ in the final state, the quoted BFs include
    $\mathcal{B}(a_0^0(980)\to K^+K^-)$. The first and second uncertainties are statistical
    and systematic, respectively.} 
 \centering
  \vspace{2mm} 
  \label{summary}
\renewcommand\arraystretch{1.2}
  \begin{tabular}{cc}
\hline\hline
      Process  &    BF (\%) \\ 
  \hline
     $D_s^+[S]\to \phi\rho^+$                                                                                                                               &$2.10\pm0.09\pm0.13$ \\
     $D_s^+[P]\to \phi\rho^+$                                                                                                                               &$0.52\pm0.05\pm0.02$ \\      
     $D_s^+[D]\to \phi\rho^+$                                                                                                                               &$0.18\pm0.04\pm0.02$ \\      
     $D_s^+\to \phi\rho^+$                                                                                                                                  &$2.75\pm0.07\pm0.15$ \\
     $D_s^+[S]\to \bar{K}^{*0}K^{*+}$                                                                                                                       &$0.88\pm0.05\pm0.03$ \\     
     $D_s^+[P]\to \bar{K}^{*0}K^{*+}$                                                                                                                       &$0.37\pm0.03\pm0.02$ \\      
     $D_s^+[D]\to \bar{K}^{*0}K^{*+}$                                                                                                                       &$0.18\pm0.03\pm0.01$ \\       
     $D_s^+\to \bar{K}^{*0}K^{*+}$                                                                                                                          &$1.25\pm0.05\pm0.06$ \\
     $D_s^+\to \bar{K}^0_1(1270)K^+$,~$\bar{K}^0_1(1270)\to K^-\rho^+$                                                                                      &$0.57\pm0.05\pm0.04$ \\      
     $D_s^+\to \bar{K}^0_1(1270)K^+$,~$\bar{K}^0_1(1270)[S]\to K^{*}\pi$                                                                                    &$0.21\pm0.04\pm0.03$ \\       
     $D_s^+\to \bar{K}^0_1(1270)K^+$,~$\bar{K}^0_1(1270)[D]\to K^{*}\pi$                                                                                    &$0.07\pm0.02\pm0.01$ \\       
     $D_s^+\to \bar{K}^0_1(1270)K^+$,~$\bar{K}^0_1(1270)\to K^{*}\pi$                                                                                       &$0.29\pm0.04\pm0.04$ \\       
     $D_s^+\to \bar{K}^0_1(1400)K^+$,~$\bar{K}^0_1(1400)\to K^{*}\pi$                                                                                       &$0.44\pm0.06\pm0.07$ \\       
     $D_s^+\to a_0^0(980) \rho^+$                                                                                                                           &$0.19\pm0.03\pm0.03$ \\        
     $D_s^+\to f_1(1420)\pi^+$,~$f_1(1420)\to K^{*\mp}K^{\pm}$                                                                                              &$0.13\pm0.02\pm0.01$ \\        
     $D_s^+\to f_1(1420)\pi^+$,~$f_1(1420)\to a_0^0(980)\pi^0$                                                                                              &$0.04\pm0.01\pm0.01$ \\    
     $D_s^+\to \eta(1475)\pi^+$,~$\eta(1475)\to a_0^0(980)\pi^0$                                                                                            &$0.07\pm0.02\pm0.02$ \\ 
\hline\hline
  \end{tabular}
\end{table}

\section*{Acknowledgements}
The BESIII collaboration thanks the staff of BEPCII and the IHEP 
computing center for their strong support. This work is supported 
in part by National Key Research and Development Program of China 
under Contracts Nos. 2020YFA0406400, 2020YFA0406300; National Natural 
Science Foundation of China (NSFC) under 
Contracts Nos. 11625523, 11635010, 11735014, 11822506, 11835012, 
11935015, 11935016, 11935018, 11961141012; the Chinese Academy of 
Sciences (CAS) Large-Scale Scientific Facility Program; Joint 
Large-Scale Scientific Facility Funds of the NSFC and CAS under 
Contracts Nos. U1732263, U1832207; CAS Key Research Program of Frontier 
Sciences under Contracts Nos. QYZDJ-SSW-SLH003, QYZDJ-SSW-SLH040; 
100 Talents Program of CAS; The Institute of Nuclear and Particle Physics and Shanghai Key Laboratory for 
Particle Physics and Cosmology; ERC under Contract No. 758462; German Research 
Foundation DFG under Contract No. 443159800 and Collaborative Research 
Center Contracts No. CRC 1044, No. FOR 2359, and No. GRK 214; Istituto Nazionale di 
Fisica Nucleare, Italy; Ministry of Development of Turkey under Contract 
No. DPT2006K-120470; National Science and Technology fund; Olle Engkvist 
Foundation under Contract No. 200-0605; STFC (United Kingdom); The Knut
 and Alice Wallenberg Foundation (Sweden) under Contract No. 2016.0157; 
The Royal Society, UK under Contracts Nos. DH140054, DH160214; The Swedish
 Research Council; U. S. Department of Energy under Contracts 
Nos. DE-FG02-05ER41374, DE-SC-0012069.
\begin{appendix}
\section{\boldmath Fixed Relations of some
Amplitudes}\label{relations} The amplitudes that are fixed by Clebsch
Gordan coefficients and charge conjugation relations in this analysis
are listed in Table~\ref{fixedre}. The amplitudes with fixed relation
share the same magnitude ($\rho$) and phase ($\phi$).
\begin{table*}[htp]
 \centering
  \caption{The fixed relations of some amplitudes.}
  \vspace{2mm} 
  \label{fixedre}
\renewcommand\arraystretch{1.2}
  \begin{tabular}{ccc}
\hline\hline
  Index   & Amplitude   &  Relation   \\
  \hline
 $A_1$   &   $D_s^+\to \bar{K}_1^0(1270)K^+$,~$\bar{K}_1^0(1270)[S]\to \bar{K}^{*0}\pi^0$         & \\
 $A_2$   &   $D_s^+\to \bar{K}_1^0(1270)K^+$,~$\bar{K}_1^0(1270)[S]\to K^{*-}\pi^+$               & \\
 $A$     &   $D_s^+\to \bar{K}_1^0(1270)K^+$,~$\bar{K}_1^0(1270)[S]\to K^{*}\pi$                  & $A_1-\sqrt{2}*A_2$\\
  \hline
 $A_1$   &   $D_s^+\to \bar{K}_1^0(1270)K^+$,~$\bar{K}_1^0(1270)[D]\to \bar{K}^{*0}\pi^0$         & \\
 $A_2$   &   $D_s^+\to \bar{K}_1^0(1270)K^+$,~$\bar{K}_1^0(1270)[D]\to K^{*-}\pi^+$               & \\
 $A$     &   $D_s^+\to \bar{K}_1^0(1270)K^+$,~$\bar{K}_1^0(1270)[D]\to K^{*}\pi$                  & $A_1-\sqrt{2}*A_2$\\
  \hline 
 $A_1$   &   $D_s^+\to \bar{K}_1^0(1400)K^+$,~$\bar{K}_1^0(1400)[S]\to \bar{K}^{*0}\pi^0$         & \\
 $A_2$   &   $D_s^+\to \bar{K}_1^0(1400)K^+$,~$\bar{K}_1^0(1400)[S]\to K^{*-}\pi^+$               & \\
 $A$     &   $D_s^+\to \bar{K}_1^0(1400)K^+$,~$\bar{K}_1^0(1400)[S]\to K^{*}\pi$                  & $A_1-\sqrt{2}*A_2$\\
  \hline
 $A_1$   & $D_s^+\to \eta(1405)\pi^+$,~$\eta(1405)\to K^{*-}K^+$                                  & \\
 $A_2$   & $D_s^+\to \eta(1405)\pi^+$,~$\eta(1405)\to K^{*+}K^-$                                  & \\
 $A$     & $D_s^+\to \eta(1405)\pi^+$,~$\eta(1405)\to K^{*\mp}K^{\pm}$                            & $A_1-A_2$\\ 
  \hline
 $A_1$   & $D_s^+\to f_1(1420)\pi^+$,~$f_1(1420)\to K^{*-}K^+$                                    & \\
 $A_2$   & $D_s^+\to f_1(1420)\pi^+$,~$f_1(1420)\to K^{*+}K^-$                                    & \\
 $A$     & $D_s^+\to f_1(1420)\pi^+$,~$f_1(1420)\to K^{*\mp}K^{\pm}$                              & $A_1-A_2$~\cite{zhaoqiang}\\ 
\hline\hline
  \end{tabular}
\end{table*}

\section{\boldmath Amplitudes Tested}\label{othertest} Other tested
amplitudes which are found to have a significance smaller than
$3\sigma$ based on the nominal fit model are listed below.
\begin{itemize} \item {\bf Cascade amplitudes} \item[-]{$D_s^+\to
\bar{K}_1^0(1270)K^+$,~$\bar{K}_1^0(1270)[D]\to K^-\rho^+$}
\item[-]{$D_s^+\to \bar{K}_1^0(1400)K^+$,~$\bar{K}_1^0(1400)[D]\to
K^{*}\pi$} \item[-]{$D_s^+\to
\bar{K}_1^0(1270)K^+$,~$\bar{K}_1^0(1270)[P]\to
\bar{K}_0^{*}(1430)\pi$} \item[-]{$D_s^+\to
\bar{K}_1^0(1400)K^+$,~$\bar{K}_1^0(1400)[S,D]\to K^-\rho^+$}
\item[-]{$D_s^+[P]\to \phi(1680)\pi^+$,~$\phi(1680)[P]\to
K^{*\mp}K^{\pm}$} \item[-]{$D_s^+\to \eta(1405)\pi^+$,~$\eta(1405)\to
K^{*\mp}K^{\pm}$} \item[-]{$D_s^+\to \eta(1475)\pi^+$,~$\eta(1475)\to
K^{*\mp}K^{\pm}$} \item[-]{$D_s^+\to \eta(1295)\pi^+$,~$\eta(1295)\to
a_0^0(980)\pi^0$} \item[-]{$D_s^+\to \eta(1405)\pi^+$,~$\eta(1405)\to
a_0^0(980)\pi^0$} \item[-]{$D_s^+\to f_1(1285)\pi^+$,~$f_1(1285)\to
a_0^0(980)\pi^0$} \item[-]{$D_s^+\to f_1(1285)\pi^+$,~$f_1(1285)\to
K^{*\mp}K^{\pm}$} \item[-]{$D_s^+\to f_1(1510)\pi^+$,~$f_1(1510)\to
K^{*\mp}K^{\pm}$} \end{itemize} \begin{itemize} \item {\bf Three-body
amplitudes} \item[-]{$D_s^+\to
\bar{K}_1^0(1270)K^+$,~$\bar{K}_1^0(1270)[P]\to (K\pi)_{{\rm S{\text
-}wave}}\pi$} \item[-]{$D_s^+[S,P,D]\to (K^-\pi^+)_{V}K^{*+}$}
\item[-]{$D_s^+[S,P,D]\to \bar{K}^{*0}(K^+\pi^0)_{V}$}
\item[-]{$D_s^+[S,P,D]\to (K^-K^+)_{V}\rho^+$}
\item[-]{$D_s^+[S,P,D]\to \phi(\pi^+\pi^0)_{V}$}
\item[-]{$D_s^+[S,P,D]\to \phi(1680)(\pi^+\pi^0)_{V}$}
\item[-]{$D_s^+\to (K^-\rho^+)_{A}[S,D]K^+$} \item[-]{$D_s^+\to
(K^{*}\pi)_{A}[S,D]K^+$} \item[-]{$D_s^+\to (K^-\rho^+)_{P}K^+$}
\item[-]{$D_s^+\to (K^-\rho^+)_{V}K^+$} \item[-]{$D_s^+\to
(K^{*\mp}K^{\pm})_{P}\pi^+$} \item[-]{$D_s^+\to
(K^{*\mp}K^{\pm})_{V}\pi^+$} \item[-]{$D_s^+[P]\to
(K^-K^+)_{S}\rho^+$} \item[-]{$D_s^+[P]\to \phi(\pi^+\pi^0)_{S}$}
\item[-]{$D_s^+[P]\to (K^-\pi^+)_{S}K^{*+}$} \item[-]{$D_s^+[P]\to
\bar{K}^{*0}(K^+\pi^0)_{S}$} \item[-]{$D_s^+[P]\to (K^-\pi^+)_{{\rm
S{\text -}wave}}K^{*+}$} \item[-]{$D_s^+[P]\to
\bar{K}^{*0}(K^+\pi^0)_{{\rm S{\text -}wave}}$} \item[-]{$D_s^+\to
\eta(1405)\pi^+$,~$\eta(1405)\to (K^{\mp}\pi^0)_{V}K^{\pm}$}
\item[-]{$D_s^+\to \eta(1475)\pi^+$,~$\eta(1475)\to
(K^{\mp}\pi^0)_{V}K^{\pm}$} \item[-]{$D_s^+\to
\eta(1405)\pi^+$,~$\eta(1405)\to (K^{\mp}\pi^0)_{{\rm S{\text
-}wave}}K^{\pm}$} \item[-]{$D_s^+\to \eta(1475)\pi^+$,~$\eta(1475)\to
(K^{\mp}\pi^0)_{{\rm S{\text -}wave}}K^{\pm}$} \end{itemize}
\begin{itemize} \item {\bf Four-body non-resonance amplitudes}
\item[-]{$D_s^+\to ((K\pi)_{{\rm S{\text -}wave}}\pi)_{A}K^+$}
\item[-]{$(K^-(\pi^+\pi^0)_{V})_{P}K^+$}
\item[-]{$(K^-(\pi^+\pi^0)_{V})_{V}K^+$} \item[-]{$D_s^+\to
((K^{\mp}\pi^0)_{V}K^{\pm})_{P}\pi^+$} \item[-]{$D_s^+\to
((K^{\mp}\pi^0)_{V}K^{\pm})_{V}\pi^+$}
\item[-]{$((K\pi)_{V}\pi)_{A}[S,D]K^+$} \item[-]{$D_s^+\to
((\pi^+\pi^0)_{V}K^-)_{A}[S,D] K^+$} \item[-]{$D_s^+[S,P,D]\to
(K^-K^+)_{V}(\pi^+\pi^0)_{V}$} \item[-]{$D_s^+[S,P,D]\to
(K^-\pi^+)_{V}(K^+\pi^0)_{V}$} \item[-]{$D_s^+\to
(K^-\pi^+)_{S}(K^+\pi^0)_{S}$} \item[-]{$D_s^+\to
(K^-K^+)_{S}(\pi^+\pi^0)_{S}$} \end{itemize} \end{appendix}

\end{document}